\documentclass[10pt, twocolumn, comsoc]{IEEEtran}

\usepackage{graphicx,epsfig}
\usepackage[noadjust]{cite}
\usepackage{mcite}
\usepackage{amsfonts,helvet}
\usepackage{fancyhdr}
\usepackage{threeparttable}
\usepackage{epsf,epsfig}
\usepackage{amsthm}
\usepackage{amsmath}
\usepackage{siunitx}
\usepackage{amssymb}
\usepackage{dsfont}
\usepackage{subfigure}
\usepackage{color}
\usepackage{microtype}
\usepackage[linesnumbered,ruled]{algorithm2e}
\SetKwFor{While}{while}{ do}{}
\usepackage{enumerate}
\usepackage{mathtools}
\usepackage{cancel}
\usepackage{comment}
\usepackage{booktabs}
\usepackage{multirow}

\newtheorem{theorem}{Theorem}

\newtheorem{corollary}{Corollary}

\newtheorem{lemma}{Lemma}

\newtheorem{proposition}{Proposition}
\newtheoremstyle{remarkbold}{\topsep}{\topsep}{\normalfont}{}{\bfseries}{.}{ }{} 
\theoremstyle{remarkbold}\newtheorem{remark}{Remark}\theoremstyle{plain}

\newtheorem{assumption}{Assumption}
\SetKwInput{KwInput}{Initialize}
\SetKwInput{KwOutput}{Output}
\SetKwProg{Fn}{Stage 1}{}{}
\SetKwProg{Fn}{Stage 2}{}{}

\usepackage{eucal}

\begin{document}

\title{Fast and Convergent Alternating GPI Precoding for Self-Backhauled Satellite-Terrestrial Networks
}

\author{Seokjun~Park, Jongsoo~Baik, Hyoungtae~Kim, Joohyun~Son, and Jinseok~Choi

\thanks{

{Seokjun Park and Jinseok Choi are with the School of Electrical and Electronic Engineering, KAIST, Daejeon 34141, South Korea (e-mail: {\texttt{sj.park@kaist.ac.kr; jinseok@kaist.ac.kr}})}.
Jongsoo Baik, Hyoungtae Kim, and Joohyun Son are with the Space R\&D Center, Hanwha Systems, South Korea (e-mail: {\texttt{jongs100@hanwha.com; hyoungtaekim@hanwha.com; joohyun.s@hanwha.com}}).
}}

\maketitle \setcounter{page}{1}

\begin{abstract}
In this paper, we propose a  precoding framework for an integrated satellite terrestrial self-backhauled (ISTS) system, where a low Earth orbit (LEO) satellite transmits a backhaul stream to an earth station (ES), and the ES forwards its payload to the base station (BS) serving the users.
We formulate a user sum spectral efficiency (SE) maximization problem subject to a self-backhaul constraint, which ensures that the user data forwarded by the BS do not exceed the data delivered to the ES.
We then reformulate the resulting non-convex precoding problem into a form amenable to generalized power iteration (GPI).
The GPI update is reinterpreted as projected preconditioned gradient ascent, revealing that its fixed step size may cause overshooting and non-monotonic behavior.
Building on this interpretation, we first develop an alternating convergent GPI (AC-GPI), in which the fixed step is replaced by an Armijo backtracking rule, and show that a single Armijo-accepted step per block already guarantees sufficient ascent. This leads to the fast alternating convergent GPI (FAC-GPI), which performs one Armijo-accepted satellite update followed by one BS update in every round. Unlike existing backtracking-based GPI methods, whose convergence has not been established, we establish finite backtracking termination, monotone ascent, convergence to stationary points, along with a global stationarity-rate guarantee.
Simulation results demonstrate stable and fast convergence as well as performance gains over existing methods.
\end{abstract}


\section{Introduction}
The rapid growth of smart device applications has dramatically increased data rate demands, driving the need for high-frequency bands such as millimeter wave, dense small-cell deployments, and advanced beamforming \cite{busari2017millimeter,rappaport2013millimeter,andrews2014what}.
While these approaches improve throughput, their reliance on dense terrestrial infrastructure and vulnerability to propagation loss and blockage make reliable wide-area coverage difficult in terrestrial networks (TNs) \cite{yang2020partially}.
This has motivated the integration of non-terrestrial networks (NTNs), including geostationary Earth orbit and low Earth orbit (LEO) satellites, with conventional TNs to provide wide-area coverage and high-capacity connectivity \cite{liu2019qos,kodheli2021satellite,giordani2020non}.
Since a wired backhaul is often unavailable in rural or isolated areas and handheld devices with limited transmit power and antenna gain cannot support direct two-way communication with the satellite, the satellite link serves as a wireless backhaul that feeds the terrestrial BS \cite{morello2006dvb,turk2019satellite}.
In integrated satellite terrestrial self-backhauled (ISTS) networks, an earth station (ES) receives the satellite backhaul and forwards the payload to a terrestrial BS serving the users, where the backhaul and access links may reuse the same spectrum for improved spectrum utilization \cite{turk2019satellite,ge20145g,hu2018joint,bonfante20195g,wang2021joint}.


Deploying multiple-input-multiple-output (MIMO) antenna arrays in ISTS networks substantially improves the spectral efficiency (SE), provided that the resulting spatial degrees of freedom are properly exploited by the precoder design. 
In co-channel ISTS networks, the simultaneous operation of satellite backhaul and terrestrial access links introduces two inter-system interference paths.
Specifically, the terrestrial BS transmission may interfere with the ES backhaul reception, as highlighted by coexistence studies in the frequency band from 3.4 to 3.6 GHz
\cite{ITU-R-S2546}, while the satellite transmission may interfere with terrestrial users within its coverage area \cite{wang2021joint,kim2025distributed,cao2025robust}, thereby making both the user sum SE and the satellite-to-ES backhaul SE dependent on the satellite and BS precoders.
Accordingly, the satellite and BS precoders need to be jointly designed to suppress inter-system interference, efficiently exploit the available spatial degrees of freedom, and improve the access-link SE while maintaining the backhaul rate.
This coupled design leads to a challenging non-convex optimization problem over often large-dimensional precoders, motivating an algorithm that provides fast and stable convergence with low computational complexity.


\subsection{Related Work}
Beamforming optimization for multi-antenna wireless systems is challenging, since the resulting problems are non-convex and the precoders need to be designed jointly to manage inter-user interference \cite{wiesel2006linear,christensen2008weighted,dahrouj2010coordinated}.
To handle such problems,  weighted minimum mean-square error (WMMSE) \cite{park2022rate} and successive convex approximation (SCA)-based methods \cite{wang2021joint} have been widely adopted, which iteratively construct and solve convex subproblems. 
These approaches can incur substantial computational complexity, particularly for large antenna systems, due to the repeated solution of such subproblems.
From a different perspective, generalized power iteration (GPI)-based algorithms interpret the first-order optimality condition of a Rayleigh-quotient form objective as a nonlinear eigenvalue problem with eigenvector dependency (NEPv) and directly find its principal eigenvector through a power iteration, without constructing convex subproblems \cite{choi2019GPI,choi2021energyIOTJ,yoo2026scalable}.

Beamforming has also been extensively studied in satellite systems. For multibeam satellite systems, \cite{joroughi2016generalized} developed generalized multicast precoding for interference mitigation, while \cite{mosquera2018distributed} considered distributed precoding for multi-gateway satellite architectures.
In addition, beamforming has been extended to integrated satellite-terrestrial networks, where \cite{lin2018joint} studied joint beamforming for secure satellite-terrestrial communications and \cite{zhang2019joint} jointly optimized beamforming and resource allocation for terrestrial-satellite cooperation.
In spectrum-sharing scenarios, interference-aware  beamforming schemes have further been investigated to manage the  interference between satellite and terrestrial transmissions \cite{cao2025interference,cao2025robust}.
Despite these advances, these studies have primarily addressed satellite transmission performance and inter-system interference management, while the coupled satellite and BS precoder design arising from self-backhauled operation remains less explored.


Self-backhauled operation has also attracted attention as an effective means of improving spectrum utilization in integrated NTN and TN systems.
Self-backhauled technology, also known as in-band wireless backhauling, allows access and backhaul links to share the same radio spectrum and infrastructure, thus eliminating the need for a dedicated wired backhaul \cite{ge20145g,hu2018joint,bonfante20195g,wang2021joint}.
The achievable-rate trade-off between massive MIMO self-backhauled small-cell deployments and direct-access macro-cell systems has been investigated \cite{bonfante20195g}.
In addition, \cite{wang2021joint} proposed a joint beamforming scheme for mmWave ISTS networks under a self-backhaul constraint coupling the access-link sum rate with the satellite-to-ES backhaul capacity.
Such coupling requires the satellite and BS precoders to be jointly optimized.

The aforementioned works have demonstrated the potential of integrated NTN and TN systems.
However, the joint design of the satellite and BS precoders in self-backhauled ISTS networks remains challenging: the two precoders are coupled through the self-backhaul constraint, and the resulting problem is nonconvex with a dimension that grows with the numbers of antennas at the satellite and the BS and the number of users.
The SCA-based framework in \cite{wang2021joint} addresses this coupled design, but it solves a sequence of convex subproblems, which incurs considerable computational complexity as the problem dimension grows.
This motivates a precoding method that attains high SE at low computational complexity. To this end, we consider the GPI method, which has shown competitive SE with low complexity in terrestrial MIMO systems with a single transmitter \cite{choi2019GPI,choi2021energyIOTJ}. Applying it to the ISTS network, however, is not a matter of simply alternating the GPI update between the satellite and BS precoders: the fixed step inherent in the conventional GPI does not ensure monotone ascent under alternating updates, and solving each precoding subproblem to inner convergence incurs unnecessary inner iterations. In this paper, we address these issues and propose a fast GPI-based precoding algorithm for the ISTS network that achieves high SE with provable convergence.

\subsection{Contributions}
The main contributions of this paper are summarized as
\begin{itemize}
    \item We formulate a user sum SE maximization problem for the ISTS network under a self-backhaul constraint, which ensures that the user data forwarded by the BS do not exceed the data delivered to the ES. With full-power transmission, the problem reduces to the joint design of the satellite and BS precoding directions on unit spheres, and the self-backhaul constraint is handled by a partial Lagrangian whose multiplier is updated by bisection. We derive the first-order optimality condition of each precoding direction as an NEPv \cite{cai:siam:18}, which allows the two precoders to be alternately updated by GPI.

    \item Since simply alternating the GPI update between the two precoders does not ensure monotone ascent, we develop  alternating convergent GPI (AC-GPI) by replacing the fixed step of the GPI update, revealed by its PPGA interpretation \cite{yoo2026scalable}, with an Armijo backtracking rule, where each precoding direction is updated until inner convergence while the other is fixed. We prove a per-step guarantee that every accepted step terminates in finitely many backtracking trials, admits an explicit lower bound, and increases the objective in proportion to the squared block gradient with explicit preconditioner constants.

    \item The per-step guarantee implies that a single Armijo-accepted step per block already suffices for monotone ascent, which motivates the fast alternating convergent GPI (FAC-GPI). In every round, the FAC-GPI performs only one Armijo-accepted satellite update followed by one BS update, so that each block immediately accounts for the latest update of the other block and the inner-loop over-solving of the AC-GPI is avoided.

    \item For both the AC-GPI and FAC-GPI, we establish monotone non-decreasing objective values and first-order stationarity of every accumulation point, and characterize a global stationarity rate over the outer rounds. Since a round of the FAC-GPI costs two GPI steps whereas a round of the AC-GPI costs $2\bar I_{\rm in}$ steps, where $\bar I_{\rm in}$ is the average number of inner steps per block, comparing the rate bounds at an equal number of GPI steps shows that the FAC-GPI admits a tighter rate.

    \item Simulations show that the proposed FAC-GPI achieves higher user sum SE than the SCA-based joint beamforming \cite{wang2021joint}, WMMSE, and RZF baselines over a wide range of BS transmit powers and antenna numbers, while a low-complexity implementation, which exploits the structure of the KKT matrices, attains the same SE with the lowest computation time. The results further confirm that the AC-GPI and FAC-GPI ascend monotonically, unlike the fixed-step alternating GPI, and that the FAC-GPI converges in less computation time than the AC-GPI, with a growing advantage as the number of BS antennas increases.
\end{itemize}

\textit{Notation}:
$\text{a}$ is a scalar, ${\bf{a}}$ is a vector and ${\bf{A}}$ is a matrix.
The superscripts $(\cdot)^{\sf T}$, $(\cdot)^{\sf H}$, and $(\cdot)^{-1}$ denote matrix transpose, Hermitian transpose, and inversion, respectively.
$\circ$ is  Hadamard product.
$\lceil \cdot \rceil$ denotes the ceiling function.
$\mathbb{E}[\cdot]$, ${\rm Tr}(\cdot)$, and $\otimes$ are  expectation operator, trace operator, and Kronecker product, respectively.
${\bf{I}}_K$ is the  $K \times K$ identity matrix.
${\bf e}_k^{N}$ is an $N \times 1$ standard basis vector whose $k$th element is one with zeros elsewhere.
${\bf 0}_{N}$ is the $N\times 1$ zero vector.

\section{System Model}
\label{sec:sys_model}
We consider the forward link of the ISTS network, where the BS with $N$
antennas serves $K$ single-antenna  users and the LEO satellite with $M$ antennas serves the ES.
The ES, equipped with a high-gain antenna, receives the satellite backhaul and forwards the user data to the BS via a wired link, so that the aperture gain of the ES and the multi-antenna gain of the BS are both exploited. Since the satellite-ES backhaul link bottlenecks the system, the access-link sum SE cannot exceed the backhaul SE received by the ES.
We consider co-channel operation in which the satellite backhaul and terrestrial access links reuse the same
frequency resources. Accordingly, the BS transmission
may interfere with the ES backhaul reception, while
the satellite backhaul signal may interfere with
terrestrial users served by the BS
\cite{wang2021joint,cao2025robust}.

Let $s\sim\mathcal{CN}(0,1)$ denote the backhaul stream, which is precoded by ${\bf f}\in\mathbb C^M$ and transmitted with the maximum satellite power $P_s$, so that the transmitted signal ${\bf x}^{\sf sat}\in\mathbb C^M$ is
\begin{align}
    \mathbf{x}^{\sf{sat}} =  \sqrt{P_s}\mathbf{f} s.
    \label{eq:analog base_sat}
\end{align}
For the access link, the BS forwards the user payload received from the ES to the $K$ users, where the stream of the $k$th user $m_k\sim\mathcal{CN}(0,1)$ is precoded by ${\bf w}_k\in\mathbb C^N$. With ${\bf W}=[{\bf w}_1,\ldots,{\bf w}_K]\in\mathbb C^{N\times K}$, the transmitted signal ${\bf x}^{\sf bs}\in\mathbb C^N$ of the BS is expressed as
\begin{align}
    {\bf{x}}^{\sf{bs}} = \sqrt{P_t} \sum_{k=1}^{K} {{\mathbf{w}}_{k}} m_{k},
    \label{eq:analog base band_signal_bs}
\end{align}
where $P_t$ is the maximum transmit power of the BS.

The received signal at the ES is formulated as
\begin{align} \label{eq:received signal_es}
    y^{\sf es} = {\bf g}^{\sf H}{\bf x}^{\sf sat} +  ({\bf h}^{\sf int})^{\sf H}{\bf x}^{\sf bs} + { n}^{\sf sat},
\end{align}
where ${\bf g} \in \mathbb {C}^{M \times 1}$ denotes the downlink channel vector between the satellite and the ES, ${\bf h}^{\sf int} \in \mathbb {C}^{N \times 1}$ denotes the inter-system interference channel vector between the BS and the ES, and ${{ n}^{\sf sat}} \backsim {\mathcal {CN} ({ 0},(\sigma^{\sf sat})^{2})}$ represents additive white Gaussian noise.
Using \eqref{eq:analog base_sat} and \eqref{eq:analog base band_signal_bs}, the received signal in \eqref{eq:received signal_es} becomes
\begin{align}
    y^{\sf es}= \sqrt{P_s}\mathbf{g}^{\sf H}\mathbf{f} s + \sqrt{P_t}(\mathbf{h}^{\sf int})^{\sf H}\sum_{k=1}^{K} {\mathbf{w}_k} m_{k} + n^{\sf sat}.
    \label{rx_sat}
\end{align}

The received signal vector at the users is
\begin{align} \label{eq:received signal_bs}
    {\bf y}^{\sf ue} =  {\bf H}^{\sf H}{\bf x}^{\sf bs} +  ({\bf G}^{\sf int})^{\sf H}{\bf x}^{\sf sat} + {\bf n}^{\sf bs},
\end{align}
where ${\bf H} \in \mathbb {C}^{N \times K}$ denotes the downlink channel matrix between the BS and users, ${{\bf G}^{\sf int}} \in \mathbb {C}^{M \times K}$ denotes an inter-system interference channel matrix between the satellite and users, and ${{\bf n}^{\sf bs}}$ is additive white Gaussian noise with zero mean and variance $(\sigma^{\sf bs})^{2}$.
The received signal at the $k$th user is
\begin{align}
    y_{{k}}^{\sf ue} \!\!=\!\!  \sqrt{P_t}\mathbf{h}_{k}^{\sf H} {\mathbf{w}_k} m_{k} \!+\! \sqrt{P_t}\mathbf{h}_{k}^{\sf H}\sum_{i\neq k}^{K} {\mathbf{w}_{i}} m_{i}
    \!+\!\! \sqrt{P_s}({\bf g}_k^{\sf int})^{\sf H} \mathbf{f} s + n_{k}^{\sf bs},\label{rx_bs}
\end{align}
where ${\bf h}_k$ is the $k$th column of $\bf H$ and ${\bf g}_k^{\sf int}$ is the $k$th column of ${\bf G}^{\sf int}$.

As shown in \eqref{rx_sat} and \eqref{rx_bs}, ${\bf f}$ affects both the backhaul signal at the ES and the interference at the users, and ${\bf W}$ affects both the desired signals at the users and the interference at the ES. We therefore jointly design ${\bf f}$ and ${\bf W}$ at the ES, which acts as a centralized precoding controller and distributes the precoders to the BS via the wired link and to the satellite via the uplink control link, following the on-ground precoding approach \cite{joroughi2016generalized,mosquera2018distributed}.
The ES estimates the satellite-to-ES and BS-to-ES channels using pilot signals, and the BS acquires the user channel state information (CSI) through uplink training or user feedback \cite{larsson2014massive,han2023fdd} and forwards it to the ES via the wired link. The satellite-to-user interference channels are dominated by LoS components, which can be computed at the ES from the satellite ephemeris and the user locations reported via the BS; residual components can be obtained from user measurements of satellite reference signals with GNSS-based delay and Doppler pre-compensation as specified for NTN \cite{3gpp38821,lin2021from5g}.
We assume that perfect CSI\footnote{Under imperfect CSIT, the proposed precoding framework can be straightforwardly extended at the SE formulation level using achievable SE lower-bound surrogates based on channel estimates and their error covariance matrices \cite{kim2025distributed, yoo2026scalable}.}
 is available at the ES to focus on the joint precoding design and convergence properties of the proposed algorithm.

\section{Beamforming Design for  NTN and TN}
In this section, assuming perfect CSI, we formulate the self-backhaul-constrained sum SE maximization problem and recast it into a GPI-friendly form.

\subsection{Sum SE Maximization Problem}
\label{subsec:perfect}
Using the received signal in \eqref{rx_sat}, the SE of the ES is 
\begin{align}
R^{\sf es}({\bf f},{{\bf W}})
\!=\!
\log_2\!\left(
\!1\!+\!
\frac{
{\bf f}^{\sf H}{\bf g}{\bf g}^{\sf H}{\bf f}
}{
\frac{P_t}{P_s}\!\sum_{i=1}^{K}
{{\bf w}_i}^{\sf H}{\bf h}^{\sf int}({\bf h}^{\sf int})^{\sf H}{{\bf w}_i}
+\frac{(\sigma^{\sf sat})^2}{P_s}
}
\right).
\label{eq:R_ES_perfect}
\end{align}
Similarly, using the received signal in \eqref{rx_bs}, the SE of the $k$th user is given by
\begin{align}
R_k^{\sf ue}({\bf f},\!{{\bf W}})
\!=\!\log_2\!\left(\!
1\!+\!\!
\frac{
{{\bf w}^{\sf H}_k}{\bf h}_k{\bf h}_k^{\sf H}{{\bf w}_k}
}{
\sum_{i\neq k}\!
{{\bf w}^{\sf H}_i}{\bf h}_k{\bf h}_k^{\sf H}{{\bf w}_i}
\!+\!\frac{P_s}{P_t}
{\bf f}^{\sf H}{\bf g}_k^{\sf int}({\bf g}_k^{\sf int})^{\sf H}{\bf f}
\!+\!\frac{(\sigma^{\sf bs})^2}{P_t}
}\!
\right).
\label{eq:R_UE_perfect}
\end{align}
Then, using \eqref{eq:R_ES_perfect} and \eqref{eq:R_UE_perfect}, we formulate an optimization problem as
\begin{align}
\underset{\left\{\mathbf f,{\mathbf W}\right\}}{\text{maximize}}
\quad
&\sum_{k=1}^{K}
R_{k}^{\sf ue}({\bf f},{{\bf W}})
\label{eq:opt_prob}\\
\text{subject to}\quad
&{\rm Tr}({{\bf W}}{{\bf W}}^{\sf H})=1,
\label{opt_prob_pow_bs}\\
&{\rm Tr}({\bf f}{\bf f}^{\sf H})=1,
\label{opt_prob_pow_es}\\
&\sum_{k=1}^{K}
R_{k}^{\sf ue}({\bf f},{{\bf W}})
\le
R^{\sf es}({\bf f},{{\bf W}}),
\label{opt_prob_backhaul}
\end{align}
where \eqref{opt_prob_pow_bs} and \eqref{opt_prob_pow_es} impose full-power transmission at the BS and satellite, respectively. Since the user sum SE tends to increase with the BS transmit power and the backhaul SE with the satellite transmit power, both are set to their maximum, while the inter-system interference is managed through the precoding directions and the self-backhaul constraint \eqref{opt_prob_backhaul}; joint power control is left for future work.
Since the BS can forward only the user payload received by the ES, \eqref{opt_prob_backhaul} ensures that the user data forwarded by the BS do not exceed the data delivered to the ES over the satellite backhaul link.
Due to the non-convexity of the resulting optimization problem, finding a globally optimal solution is intractable.
To efficiently solve this problem using GPI, we reformulate the SE expressions into a GPI-friendly form in the following subsection.

\subsection{GPI-friendly  Form}
\label{subsec:gpi_form}
Under the unit-power normalization of the BS and satellite precoders,
${{\rm Tr}}({\bf W}{\bf W}^{\sf H})=1$ and
${{\rm Tr}}({\bf f}{\bf f}^{\sf H})=1$ in \eqref{opt_prob_pow_bs} and \eqref{opt_prob_pow_es}.
To cast the problem into a GPI-friendly form, we reformulate the SE expressions into Rayleigh-quotient forms.
Let ${\bar{\bf w}} = {\rm vec}({\bf W})$ denote the vectorized precoder.
Then, $R^{\sf es}$ can be rewritten as
\begin{align}
    R^{\sf es}\left({\mathbf{f}},\Bar{\mathbf{w}}\right) =  \log_2\left(\frac{\Bar{\mathbf{w}}^{\sf H}\mathbf{U}^{\sf{sat}}\Bar{\mathbf{w}}+{\mathbf{f}}^{\sf H} \mathbf{E}^{\sf{sat}} {\mathbf{f}}}{\Bar{\mathbf{w}}^{\sf H}\mathbf{U}^{\sf{sat}}\Bar{\mathbf{w}}+{\mathbf{f}}^{\sf H} \mathbf{S}^{\sf{sat}} {\mathbf{f}}} \right),\label{eq:re_R_com_SU}
\end{align}
where ${\bf E}^{\sf sat}\!=\!{\bf g}{\bf g}^{\sf H}
+{\bf I}_{M}\frac{(\sigma^{\sf sat})^2}{P_s}$,
${\bf S}^{\sf sat}\!=\!{\bf I}_{M}\frac{(\sigma^{\sf sat})^2}{P_s}$,
and
${\bf U}^{\sf sat}\!=\!{\bf I}_{K}\otimes\frac{P_t}{P_s}{\bf h}^{\sf int}({\bf h}^{\sf int})^{\sf H}$.
Similarly, $R_{k}^{\sf ue}$ can be reformulated as
\begin{align}
    R_{k}^{\sf ue}\left(\mathbf{{ f}}, \mathbf{{\bar w}}\right) &= \log_2\left( \frac{{\mathbf{f}}^{\sf H}\mathbf{U}_{k}^{\sf{bs}}{\mathbf{f}} + \Bar{\mathbf{w}}^{\sf H}  \mathbf{E}_{k}^{\sf{bs}} \Bar{\mathbf{w}}}{{\mathbf{f}}^{\sf H}\mathbf{U}_{k}^{\sf{bs}}{\mathbf{f}} + \Bar{\mathbf{w}}^{\sf H}\mathbf{S}_{k}^{\sf{bs}}\Bar{\mathbf{w}}}\right), \label{eq:re_R_com_TU}
\end{align}
where  ${\bf E}_{k}^{\sf bs}={\bf I}_{K}\otimes{\bf h}_k{\bf h}_k^{\sf H}+{\bf I}_{NK}\frac{(\sigma^{\sf bs})^2}{P_t}$, ${\bf S}_{k}^{\sf bs}
=
{\bf E}_{k}^{\sf bs}
-
{\rm diag}({\bf e}_{k}^{K})
\otimes
{\bf h}_k{\bf h}_k^{\sf H}$, and ${\bf U}_{k}^{\sf bs}=\frac{P_s}{P_t}{\bf g}_{k}^{\sf int}({\bf g}_{k}^{\sf int})^{\sf H}.$
Accordingly, using \eqref{eq:re_R_com_SU} and \eqref{eq:re_R_com_TU}, the optimization problem in \eqref{eq:opt_prob} becomes
\begin{align}
   \underset{\left\{\mathbf{ f}, \mathbf{{\bar w}}\right\}}{\text{maximize}}
    \;\; &\sum_{k = 1}^{K} R_{k}^{\sf{ue}}\left(\mathbf{{ f}}, \mathbf{{\bar w}}\right)
    \label{eq:opt_prob_v2}\\
    \text{subject to} \quad
    &\|{\bar{{\bf w}}}\| = 1, \|{{{\bf f}}}\| = 1,
    \label{eq:constraint_power}\\
    \label{eq:constraint}
    & \sum_{k = 1}^{K} { R}_{k}^{\sf{ue}}\left(\mathbf{{ f}}, \mathbf{{\bar w}}\right) \!\leq\!  R^{\sf es}\left({\mathbf{f}},\Bar{\mathbf{w}}\right).
\end{align}
Then, we handle the self-backhaul constraint in \eqref{eq:constraint} via a Lagrange multiplier $\mu$, while the unit-norm constraints in \eqref{eq:constraint_power} are kept as the feasible set $\mathcal S_{\bf f}\times\mathcal S_{\bar{\bf w}}$, where $\mathcal S_{\bf f}=\{{\bf f}\in\mathbb C^{M}:\|{\bf f}\|_2=1\}$ and $\mathcal S_{\bar{\bf w}}=\{\bar{\bf w}\in\mathbb C^{NK}:\|\bar{\bf w}\|_2=1\}$. The resulting partial Lagrangian, restricted to this set, is
\begin{align}
    L({{\bf f}},\!\bar{{\bf w}},\mu) &=\! \sum_{k = 1}^{K}\! R_{k}^{\sf{ue}}\!\left(\mathbf{{ f}},\! \mathbf{{\bar w}}\right)
    \!+\! \mu (R^{\sf es}\left({\mathbf{f}},\!\bar{\mathbf{w}}\right) \!-\! \sum_{k = 1}^{K}\! R_{k}^{\sf{ue}}\!\left(\mathbf{{ f}}, \mathbf{{\bar w}}\right)) 
    \\
\label{eq:Lagrangian_v1}
    &= \ln{\lambda({{\bf f}},\bar{{\bf w}},\mu)}, \quad ({\bf f},\bar{\bf w})\in\mathcal S_{\bf f}\times\mathcal S_{\bar{\bf w}},
\end{align}
where $\mu$ is the Lagrange multiplier and $
    \lambda({{\bf f}},\!\bar{{\bf w}},\mu\!) \!=\!\!
\left(\!\frac{\Bar{\mathbf{w}}^{\sf H}\mathbf{U}^{\sf{sat}}\Bar{\mathbf{w}}+{\mathbf{f}}^{\sf H} \mathbf{E}^{\sf{sat}} {\mathbf{f}}}{\Bar{\mathbf{w}}^{\sf H}\mathbf{U}^{\sf{sat}}\Bar{\mathbf{w}}+{\mathbf{f}}^{\sf H} \mathbf{S}^{\sf{sat}} {\mathbf{f}} } \right)^{\frac{\mu}{\ln{2}}}\prod\limits_{k=1}^{K} \left( \frac{{\mathbf{f}}^{\sf H}\mathbf{U}_{k}^{\sf{bs}}{\mathbf{f}} + \Bar{\mathbf{w}}^{\sf H}  \mathbf{E}_{k}^{\sf{bs}} \Bar{\mathbf{w}}}{{\mathbf{f}}^{\sf H}\mathbf{U}_{k}^{\sf{bs}}{\mathbf{f}} + \Bar{\mathbf{w}}^{\sf H}\mathbf{S}_{k}^{\sf{bs}}\Bar{\mathbf{w}}}  \right)^{\frac{(1-\mu)}{\ln2}}.$

\subsection{Precoding Direction Optimization}
\label{subsec:precoding_dir}
Since the partial Lagrangian in \eqref{eq:Lagrangian_v1} is non-convex in ${\bf f}$ and $\bar{\bf w}$, we adopt an alternating optimization approach. Exploiting its Rayleigh-quotient form, we derive the first-order Karush–Kuhn–Tucker (KKT) condition of \eqref{eq:Lagrangian_v1} with respect to each precoding direction, with the other fixed, and cast it as a nonlinear eigenvalue problem with eigenvector dependency (NEPv) \cite{cai:siam:18}, as shown in Lemma~\ref{lem:1} for ${\bf f}$.
\begin{lemma} \label{lem:1}
For given ${\bar{\bf w}}$ and $\mu$, the first-order optimality condition of the partial Lagrangian in \eqref{eq:Lagrangian_v1} on $\mathcal S_{\bf f}$ is satisfied if the following holds:
\begin{align}
\mathbf{B}^{-1}_{\sf KKT}\left(\mathbf{f}\right)\mathbf{A}_{\sf KKT}\left(\mathbf{f}\right)\mathbf{f} = \lambda\left(\mathbf{f}\right) \mathbf{f},\label{KKT_sat}
\end{align}
where
\begin{align}
    &\mathbf{A}_{\sf KKT}\left(\mathbf{f}\right)
    = \lambda_{\sf num}(\mathbf{f}) \times
    \label{A(w)}
    \\
    \nonumber
    &\left\{\frac{\mu(\mathbf{E}^{\sf{sat}}+c{\bf I}_M)}{\mathbf{f}^{\sf H}(\mathbf{E}^{\sf{sat}}+c{\bf I}_M)\mathbf{f}}+ (1-\mu)\!\! \sum_{k = 1}^{K} \!\left(\!\frac{(\mathbf{U}_{k}^{\sf{bs}}+c^{\sf E}_{k}{\bf I}_M)}{\mathbf{f}^{\sf H}(\mathbf{U}_{k}^{\sf{bs}}+c^{\sf E}_{k}{\bf I}_M)\mathbf{f}}\!\!\right)\right\},
\end{align}
\begin{align}
\label{B(w)}
&\mathbf{B}_{\sf KKT}\left(\mathbf{f}\right)
    = \lambda_{\sf den}(\mathbf{f})
    \times
    \\
    \nonumber
    &\left\{\frac{\mu(\mathbf{S}^{\sf{sat}}
    +c{\bf I}_M)}{ \mathbf{f}^{\sf H}(\mathbf{S}^{\sf{sat}}+c{\bf I}_M)\mathbf{f}}+\! (1-\mu)\! \sum_{k = 1}^{K} \!\left(\!\frac{(\mathbf{U}_{k}^{\sf{bs}}+c^{\sf S}_{k}{\bf I}_M)}{\mathbf{f}^{\sf H}(\mathbf{U}_{k}^{\sf{bs}}+c^{\sf S}_{k}{\bf I}_M)\mathbf{f} }\right)\!\!\right\},
\end{align}
where $c = \bar{\bf w}^{\sf H}{\bf U}^{\sf sat}\bar{\bf w}$, $c^{\sf E}_{k} = \bar{\bf w}^{\sf H}{\bf E}^{\sf bs}_{k}\bar{\bf w}$, and $c^{\sf S}_{k} = \bar{\bf w}^{\sf H}{\bf S}^{\sf bs}_{k}\bar{\bf w}.$
Here,
$\lambda_{\sf num}(\mathbf{f})$ and $\lambda_{\sf den}(\mathbf{f})$ can be any values that satisfy  $\lambda(\mathbf{f}) = \frac{\lambda_{\sf num}(\mathbf{f})}{\lambda_{\sf den}(\mathbf{f})}$.
\begin{proof}
For fixed $\bar{\bf w}$ and $\mu$, the objective in \eqref{eq:Lagrangian_v1} can be written as the ${\bf f}$-block objective $\tilde L_{\bf f}({\bf f};\bar{\bf w},\mu):=L_1({\bf f};\bar{\bf w},\mu)+L_2({\bf f};\bar{\bf w},\mu)$ with
\begin{align}
    \label{eq:objec_f_1}
    &L_1(\mathbf{f}; \bar{{\bf w}},  \mu) = (1-\mu)\sum_{k=1}^{K}\log_2\left( \frac{\mathbf{f}^{\sf H}(\mathbf{U}_{k}^{\sf{bs}}+c^{\sf E}_{k}{\bf I}_M)\mathbf{f} }{\mathbf{f}^{\sf H}(\mathbf{U}_{k}^{\sf{bs}}+c^{\sf S}_{k}{\bf I}_M)\mathbf{f} }\right),
    \\
    \label{eq:objec_f_2}
    &L_2(\mathbf{f}; \bar{{\bf w}},  \mu) = \mu \log_2\left(\frac{\mathbf{f}^{\sf H} (\mathbf{E}^{\sf{sat}}+c{\bf I}_M) \mathbf{f}}{\mathbf{f}^{\sf H} (\mathbf{S}^{\sf{sat}}+c{\bf I}_M) \mathbf{f} } \right).
\end{align}
Although \eqref{eq:Lagrangian_v1} is defined on $\mathcal S_{\bf f}$, $\tilde L_{\bf f}$ is a sum of logarithms of ratios of quadratic forms and is thus well defined for all ${\bf f}\neq{\bf 0}$ and invariant to scaling of ${\bf f}$, i.e., $\tilde L_{\bf f}(\alpha{\bf f};\bar{\bf w},\mu)=\tilde L_{\bf f}({\bf f};\bar{\bf w},\mu)$ for all $\alpha\neq0$. Differentiating this identity with respect to $\alpha$ at $\alpha=1$ gives $\langle\partial\tilde L_{\bf f}/\partial{\bf f}^{\sf H},{\bf f}\rangle=0$, so that the Euclidean gradient of $\tilde L_{\bf f}$ is tangent to $\mathcal S_{\bf f}$. We therefore find the stationarity condition by setting this gradient to zero.
The partial derivative of $L_1$ and $L_2$ can be computed as 
\begin{align}
     \frac{\partial L_1}{\partial \mathbf{f}^{\sf H}} =\frac{(1\!-\!\mu)}{\ln{2}} \!\sum_{k=1}^{K}\!\left(\!\frac{\!(\mathbf{U}_{k}^{\sf{bs}}+c^{\sf E}_{k}{\bf I}_M)\mathbf{f}}{\mathbf{f}^{\sf H}\!(\mathbf{U}_{k}^{\sf{bs}}\!+\!c^{\sf E}_{k}{\bf I}_M)\mathbf{f}} \!-\! \frac{(\mathbf{U}_{k}^{\sf{bs}}+c^{\sf S}_{k}{\bf I}_M)\mathbf{f}}{\mathbf{f}^{\sf H}(\mathbf{U}_{k}^{\sf{bs}}\!+\!c^{\sf S}_{k}{\bf I}_M)\mathbf{f} }\!\right)\!,
\end{align}
\begin{align}
    & \frac{\partial L_2}{\partial \mathbf{f}^{\sf H}} \!=\!\frac{\mu}{\ln{2}}\!\left(\!\frac{(\mathbf{E}^{\sf{sat}}+c{\bf I}_M)\mathbf{f}}{\mathbf{f}^{\sf H}(\mathbf{E}^{\sf{sat}}\!+\!c{\bf I}_M)\mathbf{f}} \!- \frac{(\mathbf{S}^{\sf{sat}}
    +c{\bf I}_M)\mathbf{f}}{ \mathbf{f}^{\sf H}(\mathbf{S}^{\sf{sat}}\!+\!c{\bf I}_M)\mathbf{f}}\! \right)\!.
\end{align}
Subsequently, the first-order optimality condition is given as $\frac{\partial \tilde L_{\bf f}}{\partial \mathbf{f}^{\sf H}} = \frac{\partial L_1}{\partial \mathbf{f}^{\sf H}} + \frac{\partial L_2}{\partial \mathbf{f}^{\sf H}} = {\bf 0}$.
This can be rearranged into the form shown below:
\begin{align}
    {\bf{A}}_{\sf KKT}(\mathbf{f})  {\mathbf{f}} = \lambda(\mathbf{f}) {\bf{B}}_{\sf KKT} (\mathbf{f}) \mathbf{f}.
\end{align}
Since the gradient of $\tilde L_{\bf f}$ is tangent to $\mathcal S_{\bf f}$, the KKT condition $\partial\tilde L_{\bf f}/\partial{\bf f}^{\sf H}=\nu{\bf f}$ for the constraint $\|{\bf f}\|_2=1$ forces $\nu=0$; hence setting the gradient to zero is equivalent to first-order stationarity of \eqref{eq:Lagrangian_v1} on $\mathcal S_{\bf f}$. In this regard, the condition in \eqref{KKT_sat} is obtained, which completes the proof.
\end{proof}
\end{lemma}

Considering ${\bf f}$ as an eigenvector with the corresponding eigenvalue $\lambda({\bf f})$ of  ${\bf{B}}_{\sf KKT}^{-1} ({\bf f}){\bf{A}}_{\sf KKT}({\bf f})$, the KKT stationarity condition in \eqref{KKT_sat} can be interpreted as a functional eigenvalue problem.
Since $\lambda({\bf f})$ is the objective in \eqref{eq:Lagrangian_v1} up to a logarithm, the leading eigenvector of \eqref{KKT_sat} corresponds to the stationary point with the largest objective value among those satisfying \eqref{KKT_sat}, and can be efficiently computed by GPI \cite{choi2019GPI,park2025optimizing,park2022rate}.
We summarize this observation as
\begin{proposition}
    \label{prof:main} Denoting the leading eigenvector for the problem \eqref{KKT_sat} to be ${\bf f}^{\star }$ and its corresponding eigenvalue to be $\lambda^\star({\bf f}^{\star}) $, i.e., ${\bf{B}}_{\sf KKT}^{-1}({\bf f}^{\star}){\bf{A}}_{\sf KKT}({\bf f}^{\star}) {\bf f}^{\star} = \lambda^{\star}({\bf f}^{\star})  {\bf f}^{\star}$, the eigenvector ${\bf f}^{\star}$ is the stationary point that is likely to be the superior locally optimal solution of the problem in \eqref{eq:Lagrangian_v1}.
\end{proposition}
Let ${\bf f}_t$ denote the  precoding vector at  $t$th iteration. Using \eqref{A(w)} and \eqref{B(w)}, we first construct the matrices ${\bf A}_{\sf KKT}$ and ${\bf B}_{\sf KKT}$, and subsequently update ${\bf f}_{t+1}$ as follows:
\begin{align}
    {\bf f}_{t+1} = \frac{{\bf{B}}_{\sf KKT}^{-1} ({\bf f}_{t}) {\bf{A}}_{\sf KKT} ({\bf f}_{t}) {\bf f}_{t}}{\| {\bf{B}}_{\sf KKT}^{-1} ({\bf f}_{t}) {\bf{A}}_{\sf KKT} ({\bf f}_{t}) {\bf f}_{t} \|}.
    \label{eq:KKTinver}
\end{align}
The algorithm repeats the updating procedure until convergence, i.e., until the change $\|{\bf f}_{t+1}-{\bf f}_t\|_2$ falls below a tolerance $\epsilon_{\rm in}>0$.

Subsequently, we can derive the first-order optimality condition of \eqref{eq:Lagrangian_v1} with respect to
$\bar{\bf{w}}$, assuming that ${\bf f}$ is given.
\begin{lemma} \label{lem:2}
For given ${\bf f}$ and $\mu$, the first-order optimality condition of the partial Lagrangian in \eqref{eq:Lagrangian_v1} on $\mathcal S_{\bar{\bf w}}$ is satisfied if the following holds:
    \begin{align}
    \mathbf{D}_{\sf KKT}^{-1}\left(\Bar{\mathbf{w}}\right)\mathbf{C}_{\sf KKT}\left(\Bar{\mathbf{w}}\right)\Bar{\mathbf{w}} = \lambda\left(\Bar{\mathbf{w}}\right) \Bar{\mathbf{w}},\label{KKT_bs}
\end{align}
where
\begin{align}
    \label{eq:CKKT}
    &\mathbf{C}_{\sf KKT}\left(\Bar{\mathbf{w}}\right) = \lambda_{\sf num}\left(\Bar{\mathbf{w}}\right) \times
    \\
        \nonumber
    &  \left\{\frac{\mu(\mathbf{U}^{\sf{sat}}+d^{\sf E}\mathbf{I}_{NK})}{\bar{\mathbf{w}}^{\sf H}(\mathbf{U}^{\sf{sat}}+d^{\sf E}\mathbf{I}_{NK})\bar{\mathbf{w}}}\!+ (1-\mu)\!\! \sum_{k = 1}^{K}\! \left(\frac{(\mathbf{E}_{k}^{\sf{bs}}+d_{k}\mathbf{I}_{NK})}{\bar{\mathbf{w}}^{\sf H}(\mathbf{E}_{k}^{\sf{bs}}+d_{k}\mathbf{I}_{NK})\bar{\mathbf{w}}}\right)\!\!\right\},
\end{align}
\begin{align}
\label{eq:DKKT}
&\mathbf{D}_{\sf KKT}\!\left(\Bar{\mathbf{w}}\right)
= \lambda_{\sf den}(\Bar{\mathbf{w}})\times
\\
\nonumber
&\left\{\,
\frac{\mu(\mathbf{U}^{\sf{sat}}+d^{\sf S}\mathbf{I}_{NK})}{\bar{\mathbf{w}}^{\sf H}(\mathbf{U}^{\sf{sat}}+d^{\sf S}\mathbf{I}_{NK})\bar{\mathbf{w}}}\!+\!\! (1-\mu)\!\! \sum_{k = 1}^{K}
\left(
\frac{(\mathbf{S}_{k}^{\sf{bs}}+d_{k}\mathbf{I}_{NK})}{\bar{\mathbf{w}}^{\sf H}(\mathbf{S}_{k}^{\sf{bs}}+d_{k}\mathbf{I}_{NK})\bar{\mathbf{w}}}
\!\right)
\!\!\right\},
\end{align}
where $d^{\sf E} = {\bf f}^{\sf H}{\bf E}^{\sf sat}{\bf f},
d^{\sf S} = {\bf f}^{\sf H}{\bf S}^{\sf sat}{\bf f},$ and $
d_{k} = {\bf f}^{\sf H}{\bf U}^{\sf bs}_{k}{\bf f}.$
Here, $\lambda_{\sf num}(\bar {\bf{w}})$ and $\lambda_{\sf den}(\bar {\bf{w}})$ can be any values that satisfy  $\lambda(\bar{{\bf w}}) = \frac{\lambda_{\sf num}(\bar{{\bf w}})}{\lambda_{\sf den}(\bar{{\bf w}})}$.
\begin{proof}
For fixed ${\bf f}$ and $\mu$, the objective in \eqref{eq:Lagrangian_v1} can be written as the $\bar{\bf w}$-block objective $\tilde L_{\bar{\bf w}}(\bar{\bf w};{\bf f},\mu):=L_1(\bar{\bf w};{\bf f},\mu)+L_2(\bar{\bf w};{\bf f},\mu)$ with
\begin{align}
    \label{eq:obj_w_1}
    &L_1(\bar{{\bf w}}; {\bf f},  \mu) \!=\! (1-\mu)\sum_{k=1}^{K}\log_2\left( \frac{ \Bar{\mathbf{w}}^{\sf H}  (\mathbf{E}_{k}^{\sf{bs}}+d_k{\bf I}_{NK}) \Bar{\mathbf{w}}}{\Bar{\mathbf{w}}^{\sf H}(\mathbf{S}_{k}^{\sf{bs}}
    +d_k{\bf I}_{NK})\Bar{\mathbf{w}}}\right),
    \\
    \label{eq:obj_w_2}
    &L_2(\bar{{\bf w}}; {\bf f},  \mu)= \mu \log_2\left(\frac{\Bar{\mathbf{w}}^{\sf H}(\mathbf{U}^{\sf{sat}}+d^{\sf E}{\bf I}_{NK})\Bar{\mathbf{w}}}{\Bar{\mathbf{w}}^{\sf H}(\mathbf{U}^{\sf{sat}}+d^{\sf S}{\bf I}_{NK})\Bar{\mathbf{w}}} \right).
\end{align}
The proof follows the same steps as that of Lemma~\ref{lem:1} with the roles of ${\bf f}$ and $\bar{\bf w}$ exchanged, which yields the condition in \eqref{KKT_bs}.
\end{proof}
\end{lemma}

Similarly, the  precoder ${\bar{\mathbf w}}$ can be obtained in the same manner as for deriving ${{\mathbf f}}$, i.e., using \eqref{eq:CKKT} and \eqref{eq:DKKT}, $\bar{\bf w}_{t+1}$ is updated as
\begin{align}
    \bar{\bf w}_{t+1} = \frac{{\bf{D}}_{\sf KKT}^{-1} (\bar{\bf w}_{t}) {\bf{C}}_{\sf KKT} (\bar{\bf w}_{t}) \bar{\bf w}_{t}}{\| {\bf{D}}_{\sf KKT}^{-1} (\bar{\bf w}_{t}) {\bf{C}}_{\sf KKT} (\bar{\bf w}_{t}) \bar{\bf w}_{t} \|}.
    \label{eq:KKTinver_w}
\end{align}
The proposed algorithm iteratively updates ${\bf f}$ and $\Bar{\mathbf{w}}$ through two stages: in the first stage, ${\bf f}$ is derived while fixing $\Bar{\mathbf{w}}$, and in the second stage, $\Bar{\mathbf{w}}$ is derived while fixing ${\bf f}$.
In each stage, the GPI update in \eqref{eq:KKTinver} or \eqref{eq:KKTinver_w} is repeated until inner convergence, i.e., until the precoder change falls below $\epsilon_{\rm in}$, and the two stages are alternated until convergence.
The GPI update may overshoot and fail to monotonically increase \eqref{eq:Lagrangian_v1}, motivating the development of convergent variants in the next section. Throughout this subsection and Sections~\ref{sec:ppga}-\ref{subsec:fac}, the multiplier $\mu$ is fixed.
Its update by bisection to satisfy the self-backhaul constraint \eqref{eq:constraint} is described in Section~\ref{subsec:backhaul_control}.

\section{Fast Alternating Convergent GPI}
In this section, we interpret the GPI update as PPGA with a fixed step and replace the step by Armijo backtracking. We first present the AC-GPI, which updates each block until inner convergence, and then, based on a per-step ascent guarantee, the FAC-GPI, which performs a single Armijo-accepted step per block in every round. We then present the overall algorithm with the multiplier update and analyze the computational complexity.
\subsection{PPGA Interpretation of the GPI Update}
\label{sec:ppga}
Recall from the proofs of Lemmas~\ref{lem:1} and \ref{lem:2} that, for fixed $\bar{\bf w}$, the ${\bf f}$-block objective $\tilde L_{\bf f}$ in \eqref{eq:objec_f_1}-\eqref{eq:objec_f_2} is defined for all ${\bf f}\neq{\bf 0}$, is invariant to nonzero scaling of ${\bf f}$, and coincides with $L$ in \eqref{eq:Lagrangian_v1} on $\mathcal S_{\bf f}$; the same holds for the $\bar{\bf w}$-block objective $\tilde L_{\bar{\bf w}}$ in \eqref{eq:obj_w_1}-\eqref{eq:obj_w_2} on $\mathcal S_{\bar{\bf w}}$.
Accordingly, the projection step enforces the unit-norm constraints after each preconditioned update without changing the value of the block objective.

Let $\tilde{\bf A}({\bf f})$ and $\tilde{\bf B}({\bf f})$ denote the  matrices in \eqref{A(w)} and \eqref{B(w)}, respectively, after removing the scalar factors $\lambda_{\sf num}({\bf f})$ and $\lambda_{\sf den}({\bf f})$.
Similarly, let $\tilde{\bf C}(\bar{\bf w})$ and $\tilde{\bf D}(\bar{\bf w})$ denote the  matrices in \eqref{eq:CKKT} and \eqref{eq:DKKT}.
A standard complex matrix derivative yields $\nabla_{\bf f}\tilde L_{\bf f}
=
2\big(\tilde{\bf A}({\bf f})-\tilde{\bf B}({\bf f})\big){\bf f}$ and $
\nabla_{\bar{\bf w}}\tilde L_{\bar{\bf w}}
=
2\big(\tilde{\bf C}(\bar{\bf w})-\tilde{\bf D}(\bar{\bf w})\big)\bar{\bf w}.$
With the GPI-induced preconditioners ${\bf P}_{\bf f}({\bf f})=\tilde{\bf B}({\bf f})^{-1}$ and ${\bf P}_{\bar{\bf w}}(\bar{\bf w})=\tilde{\bf D}(\bar{\bf w})^{-1}$, the GPI update can be written as a PPGA step.
For the $\mathbf f$-block, the GPI mapping before normalization can be expressed \cite{yoo2026scalable},
\begin{align}
\tilde{\bf B}^{-1}\tilde{\bf A}{\bf f}
=\tilde{\bf B}^{-1}\left[\tilde{\bf B}+(\tilde{\bf A}-\tilde{\bf B})\right]{{\bf f}}=
{\bf f}
+\frac{1}{2}{\bf P}_{\bf f}\nabla_{\bf f}\tilde L_{\bf f},
\label{eq:gpi_ppga}
\end{align}
and the same identity holds for the $\bar{\bf w}$-block.
Hence, the GPI update is a PPGA step with the fixed step $\eta=1/2$ followed by projection onto the unit sphere. Since each updated block feeds the subsequent block update, this fixed step may propagate an overshooting update and cause oscillation, which motivates the Armijo backtracking rule introduced next.

\subsection{Alternating Convergent GPI}
\label{subsec:armijo}
We now replace the fixed step by a backtracked step that guarantees sufficient ascent, and develop the alternating convergent GPI (AC-GPI).
We impose the following assumptions throughout the subsequent analysis unless stated otherwise.
\begin{assumption}[Euclidean smoothness near the spheres]
\label{ass:smooth_ppga}
There exists an open neighborhood $\mathcal{U}$ containing
$\mathcal{S}_{\bf f}\times\mathcal{S}_{\bar{\bf w}}$ such that the Euclidean gradient of each block objective, viewed as a function of ${\bf x}=({\bf f},\bar{\bf w})$, is $L_g$-Lipschitz continuous on $\mathcal{U}$:
\begin{align}
\|\nabla \tilde L({\bf x})-\nabla \tilde L({\bf y})\|_2
\le
L_g\|{\bf x}-{\bf y}\|_2,
\forall\,{\bf x},{\bf y}\in\mathcal{U}.
\end{align}
\end{assumption}
\begin{assumption}[Uniformly bounded preconditioners]
\label{ass:precond_ppga}
There exist constants $0<m_{\bf f}\le M_{\bf f}<\infty$ and
$0<m_{\bar{\bf w}}\le M_{\bar{\bf w}}<\infty$ (given explicitly in Remark~\ref{rem:precond_const}) such that, for all
$({\bf f},\bar{\bf w})\in\mathcal{S}_{\bf f}\times\mathcal{S}_{\bar{\bf w}}$,
\begin{align}
m_{\bf f}{\bf I}\preceq {\bf P}_{\bf f}({\bf f})\preceq M_{\bf f}{\bf I},
m_{\bar{\bf w}}{\bf I}\preceq {\bf P}_{\bar{\bf w}}(\bar{\bf w})\preceq M_{\bar{\bf w}}{\bf I}.
\label{eq:bounding_ppga}
\end{align}
\end{assumption}
\begin{remark}[Explicit preconditioner bounds]
\label{rem:precond_const}
Assumption~\ref{ass:precond_ppga} always holds in the considered system, with explicit constants. On $\mathcal S_{\bf f}$, $\tilde{\bf B}({\bf f})=\mu\,{\bf A}_0/({\bf f}^{\sf H}{\bf A}_0{\bf f})+(1-\mu)\sum_{k=1}^K{\bf A}_k/({\bf f}^{\sf H}{\bf A}_k{\bf f})$ with ${\bf A}_0={\bf S}^{\sf sat}+c{\bf I}_M\succ{\bf 0}$ and ${\bf A}_k={\bf U}_k^{\sf bs}+c_k^{\sf S}{\bf I}_M\succ{\bf 0}$. For any ${\bf A}\succ{\bf 0}$ and $\|{\bf f}\|_2=1$, the eigenvalues of ${\bf A}/({\bf f}^{\sf H}{\bf A}{\bf f})$ lie in $[\kappa({\bf A})^{-1},\kappa({\bf A})]$, where $\kappa(\cdot)$ denotes the condition number. Since ${\bf S}^{\sf sat}$ is a scaled identity, $\kappa({\bf A}_0)=1$, and since ${\bf U}_k^{\sf bs}$ has rank one and $c_k^{\sf S}=\bar{\bf w}^{\sf H}{\bf S}_k^{\sf bs}\bar{\bf w}\ge(\sigma^{\sf bs})^2/P_t$ on $\mathcal S_{\bar{\bf w}}$, we have $\kappa({\bf A}_k)\le\kappa_k:=1+P_s\|{\bf g}_k^{\sf int}\|_2^2/(\sigma^{\sf bs})^2$. Hence, \eqref{eq:bounding_ppga} holds with
\begin{align}
\textstyle m_{\bf f}\!=\!\big(\mu\!+\!(1\!-\!\mu)\sum_{k=1}^K\kappa_k\big)^{-1}\!,\ \,
M_{\bf f}\!=\!\big(\mu\!+\!(1\!-\!\mu)\sum_{k=1}^K\kappa_k^{-1}\big)^{-1}\!.
\nonumber
\end{align}
Similarly, for $\tilde{\bf D}(\bar{\bf w})$, using $d^{\sf S}=(\sigma^{\sf sat})^2/P_s$ on $\mathcal S_{\bf f}$ and ${\bf S}_k^{\sf bs}\succeq((\sigma^{\sf bs})^2/P_t){\bf I}_{NK}$, the corresponding condition numbers are bounded by $\bar\kappa_0:=1+P_t\|{\bf h}^{\sf int}\|_2^2/(\sigma^{\sf sat})^2$ and $\bar\kappa_k:=1+P_t\|{\bf h}_k\|_2^2/(\sigma^{\sf bs})^2$, so that \eqref{eq:bounding_ppga} holds with
\begin{align}
\textstyle m_{\bar{\bf w}}\!=\!\!\big(\mu\bar\kappa_0\!+\!(1\!-\!\mu)\sum_{k=1}^K\bar\kappa_k\big)^{-1}\!,
M_{\bar{\bf w}}\!=\!\big(\mu\bar\kappa_0^{-1}\!+\!(1\!-\!\mu)\sum_{k=1}^K\bar\kappa_k^{-1}\big)^{-1}\!.
\nonumber
\end{align}
All constants are finite and positive whenever the noise variances are positive, and they depend only on the channel gains and SNRs.
\end{remark}
Based on the PPGA interpretation in \eqref{eq:gpi_ppga}, we generalize the GPI update by introducing a step size $\eta>0$.
Then, for a given ${\bf f}_t$, the trial point of the $\mathbf f$-block GPI update with step size $\eta$ is defined as
\begin{align}
{\bf z}_{{\bf f}_{t}}(\eta)
=
\mathrm{Proj}_{\mathcal{S}_{\bf f}}\!\big(
{\bf f}_{t}
+\eta\,{\bf P}_{\bf f}({\bf f}_{t})\nabla_{\bf f}\tilde L_{\bf f}({\bf f}_{t})
\big),
\label{eq:zf_def_ppga}
\end{align}
where $\mathrm{Proj}_{\mathcal{S}}({\bf z})={\bf z}/\|{\bf z}\|_2$ denotes the projection onto the unit sphere.
That is, ${\bf f}_t$ is moved along the preconditioned gradient direction ${\bf P}_{\bf f}({\bf f}_{t})\nabla_{\bf f}\tilde L_{\bf f}({\bf f}_{t})$ with step size $\eta$ and then normalized.
In particular, setting $\eta=1/2$ recovers the GPI update in \eqref{eq:KKTinver} by \eqref{eq:gpi_ppga}, whereas $\eta$ is adaptively chosen in the proposed method.
With a backtracking approach, we choose $\eta$ from $\{\eta_0,\beta\eta_0,\beta^2\eta_0,\ldots\}$ with $\eta_0>0$ and $\beta\in(0,1)$ as the first value satisfying
\begin{align}
&\tilde L_{\bf f}\!\big({\bf z}_{{\bf f}_{t}}(\eta)\big)
\ge
\tilde L_{\bf f}({\bf f}_{t})
+\sigma\eta
\big\langle \nabla_{\bf f}\tilde L_{\bf f}({\bf f}_{t}),
{\bf P}_{\bf f}({\bf f}_{t})\nabla_{\bf f}\tilde L_{\bf f}({\bf f}_{t})\big\rangle,
\label{eq:armijo_f_ppga}
\end{align}
where $\sigma\in(0,1)$ and $\langle{\bf a},{\bf b}\rangle=\operatorname{Re}\{{\bf a}^{\sf H}{\bf b}\}$ denotes the real inner product.
Similarly, with the updated satellite precoder ${\bf f}_{t+1}$ fixed, the trial
point for the $\bar{\bf w}$-block is defined as
\begin{align}
\label{eq:zr_def_ppga}
{\bf z}_{\bar{\bf w}_{t}}(\eta)
=
\mathrm{Proj}_{\mathcal{S}_{\bar{\bf w}}}\!\big(
\bar{\bf w}_{t}
+\eta\,{\bf P}_{\bar{\bf w}}(\bar{\bf w}_{t})\nabla_{\bar{\bf w}}\tilde L_{\bar{\bf w}}(\bar{\bf w}_{t})
\big),
\end{align}
and $\eta$ is accepted if
\begin{align}
&\tilde L_{\bar{\bf w}}\!\big({\bf z}_{\bar{\bf w}_{t}}\!(\!\eta)\!\big)
\!\!\ge\!
\tilde L_{\bar{\bf w}}(\bar{\bf w}_{t})
\!+\!\sigma\eta
\big\langle\! \nabla_{\bar{\bf w}}\tilde L_{\bar{\bf w}}(\bar{\bf w}_{t}\!),
{\bf P}_{\bar{\bf w}}(\bar{\bf w}_{t})\nabla_{\bar{\bf w}}\tilde L_{\bar{\bf w}}(\bar{\bf w}_{t}\!)\!\big\rangle.
\label{eq:armijo_w_ppga}
\end{align}

\begin{algorithm}[t]
\caption{AC-GPI}\label{alg:Precoding_conv}
{\bf{initialize}}: $\mathbf f_0$, $\bar{\mathbf w}_0$, and $t$

\While{$\delta_t>\epsilon$}{
  Set $\mathbf f^{(0)}\leftarrow\mathbf f_t$, and $s\leftarrow0$

  \While{$\delta^{\sf in}>\epsilon_{\rm in}$}{
    Build $\tilde{\mathbf A}(\mathbf f^{(s)})$ and $\tilde{\mathbf B}(\mathbf f^{(s)})$ from \eqref{A(w)} and \eqref{B(w)}

    Choose $\eta\in\{\eta_0,\beta\eta_0,\beta^2\eta_0,\ldots\}$ satisfying \eqref{eq:armijo_f_ppga}

    Compute $\mathbf f^{(s+1)} = \mathbf f^{(s)}+\eta\,\tilde{\mathbf B}^{-1}(\mathbf f^{(s)})\nabla_{\mathbf f}\tilde L_{\mathbf f}(\mathbf f^{(s)})$

    Normalize $\mathbf f^{(s+1)}\leftarrow\mathbf f^{(s+1)}/\|\mathbf f^{(s+1)}\|$

    $\delta^{\sf in}\leftarrow\|\mathbf f^{(s+1)}-\mathbf f^{(s)}\|$ and $s\leftarrow s+1$
  }
  Set $\mathbf f_{t+1}\leftarrow\mathbf f^{(s)}$

  Set $\bar{\mathbf w}^{(0)}\leftarrow\bar{\mathbf w}_t$, and $s\leftarrow0$

  \While{$\delta^{\sf in}>\epsilon_{\rm in}$}{
    Build $\tilde{\mathbf C}(\bar{\mathbf w}^{(s)})$ and $\tilde{\mathbf D}(\bar{\mathbf w}^{(s)})$ from \eqref{eq:CKKT} and \eqref{eq:DKKT}

    Choose $\eta\in\{\eta_0,\beta\eta_0,\beta^2\eta_0,\ldots\}$ satisfying \eqref{eq:armijo_w_ppga}

    Compute $\bar{\mathbf w}^{(s+1)} \!=\! \bar{\mathbf w}^{(s)}\!+\!\eta\,\tilde{\mathbf D}^{-1}(\!\bar{\mathbf w}^{(s)}\!)\nabla_{\bar{\mathbf w}}\tilde L_{\bar{\mathbf w}}(\bar{\mathbf w}^{(s)})$

    Normalize $\bar{\mathbf w}^{(s+1)}\leftarrow\bar{\mathbf w}^{(s+1)}/\|\bar{\mathbf w}^{(s+1)}\|$

    $\delta^{\sf in}\leftarrow\|\bar{\mathbf w}^{(s+1)}-\bar{\mathbf w}^{(s)}\|$ and $s\leftarrow s+1$
  }
  Set $\bar{\mathbf w}_{t+1}\leftarrow\bar{\mathbf w}^{(s)}$

  $\delta_t\leftarrow\|\mathbf f_{t+1}-\mathbf f_t\| + \|\bar{\mathbf w}_{t+1}-\bar{\mathbf w}_t\|$

  $t\leftarrow t+1$
}
\Return $\mathbf f_t$, $\bar{\mathbf w}_t$
\end{algorithm}
Using the Armijo-backtracked GPI step, we obtain the AC-GPI, which retains the block-wise structure of the alternating GPI, i.e., each precoding direction is updated until inner convergence while the other is fixed, but replaces each GPI update with the Armijo-accepted PPGA update, as summarized in Algorithm~\ref{alg:Precoding_conv}. 

Since every accepted step satisfies the Armijo condition and the block objectives coincide with \eqref{eq:Lagrangian_v1} on the corresponding unit spheres, each inner update of the AC-GPI yields sufficient ascent of \eqref{eq:Lagrangian_v1}, as summarized in the following per-step guarantee.
\begin{lemma}[Per-step Armijo guarantee]
\label{lem:bt_term_ppga}
Let $\bar{\bf w}_t$ and $\mu$ be fixed.
If $\nabla_{\bf f}\tilde L_{\bf f}({\bf f}_t)={\bf 0}$, the first trial step satisfies \eqref{eq:armijo_f_ppga} and ${\bf f}_t$ remains unchanged.
Otherwise, the Armijo rule in \eqref{eq:armijo_f_ppga} terminates in finitely many backtracking steps, the accepted step satisfies $\eta_{{\bf f}_t}\ge\eta_{\bf f}=\min\{\eta_0,\beta\bar\eta_{\bf f}\}$, where $\bar\eta_{\bf f}=2(1-\sigma)m_{\bf f}/(L_gM_{\bf f}^2)$, and the accepted update ${\bf f}_{t+1}={\bf z}_{{\bf f}_t}(\eta_{{\bf f}_t})$ satisfies
\begin{align}
L({\bf f}_{t+1},\bar{\bf w}_{t})-L({\bf f}_{t},\bar{\bf w}_{t})
\ge
\sigma\eta_{\bf f}m_{\bf f}
\|\nabla_{\bf f}\tilde L_{\bf f}({\bf f}_t)\|_2^2.
\label{eq:armijo_ascent}
\end{align}
The same holds for the $\bar{\bf w}$-block with \eqref{eq:zr_def_ppga} and \eqref{eq:armijo_w_ppga} for fixed ${\bf f}_{t+1}$, where $\eta_{\bar{\bf w}}=\min\{\eta_0,\beta\bar\eta_{\bar{\bf w}}\}$ and $\bar\eta_{\bar{\bf w}}=2(1-\sigma)m_{\bar{\bf w}}/(L_gM_{\bar{\bf w}}^2)$.
\end{lemma}
\begin{proof}
Let
${\bf g}_{{\bf f}_t}=\nabla_{\bf f}\tilde L_{\bf f}({\bf f}_t)$ and
${\bf d}_{{\bf f}_t}={\bf P}_{\bf f}({\bf f}_t){\bf g}_{{\bf f}_t}$.
If ${\bf g}_{{\bf f}_t}={\bf 0}$, then ${\bf z}_{{\bf f}_t}(\eta_0)={\bf f}_t$, and \eqref{eq:armijo_f_ppga} holds with equality.
We thus assume ${\bf g}_{{\bf f}_t}\neq{\bf 0}$.
Define $\phi(\eta)=\tilde L_{\bf f}({\bf z}_{{\bf f}_{t}}(\eta))$.
For sufficiently small $\eta$ such that ${\bf f}_t+\eta{\bf d}_{{\bf f}_t}\neq {\bf 0}$, $\phi(\eta)=\tilde L_{\bf f}({\bf f}_t+\eta{\bf d}_{{\bf f}_t}).$
Thus,
\begin{align}
\phi'(0)
=
\big\langle{\bf g}_{{\bf f}_t},{\bf P}_{\bf f}({\bf f}_t){\bf g}_{{\bf f}_t}\big\rangle
\ge
m_{\bf f}\|{\bf g}_{{\bf f}_t}\|_2^2.
\end{align}
If ${\bf g}_{{\bf f}_t}\neq{\bf 0}$, then $\phi'(0)>0$.
For $\eta\le\eta_0$ with $\eta_0$ chosen such that the segment $\{{\bf f}_t+\tau{\bf d}_{{\bf f}_t}:\tau\in[0,\eta]\}$ lies in $\mathcal U$ and is bounded away from the origin, using the smoothness assumption \cite{nocedal2006numerical} (the lower bound of the descent lemma applied to the extension $\tilde L_{\bf f}$ along this segment),
\begin{align}
\phi(\eta)
\ge
\phi(0)+\eta\phi'(0)-\frac{L_g}{2}\eta^2\|{\bf d}_{{\bf f}_t}\|_2^2.
\end{align}
Accordingly, the Armijo condition in \eqref{eq:armijo_f_ppga}  is satisfied whenever
\begin{align}
0<\eta
\le
\frac{2(1-\sigma)\phi'(0)}{L_g\|{\bf d}_{{\bf f}_t}\|_2^2}.
\label{eq:armijo_suff}
\end{align}
Since $\phi'(0)\ge m_{\bf f}\|{\bf g}_{{\bf f}_t}\|_2^2$ and
$\|{\bf d}_{{\bf f}_t}\|_2\le M_{\bf f}\|{\bf g}_{{\bf f}_t}\|_2$, the right-hand side of \eqref{eq:armijo_suff} is lower bounded as
\begin{align}
\frac{2(1-\sigma)\phi'(0)}{L_g\|{\bf d}_{{\bf f}_t}\|_2^2} \ge
\frac{2(1-\sigma)m_{\bf f}}{L_gM_{\bf f}^2} =\bar\eta_{{\bf f}},
\end{align}  
so that every $\eta\le\bar\eta_{{\bf f}}$ satisfies \eqref{eq:armijo_f_ppga}.
The lower bound on the accepted step follows because the first accepted backtracking step is at least $\min\{\eta_0,\beta\bar\eta_{\bf f}\}$ whenever ${\bf g}_{{\bf f}_t}\neq {\bf 0}$.
Furthermore, combining \eqref{eq:armijo_f_ppga} with $\phi'(0)\ge m_{\bf f}\|{\bf g}_{{\bf f}_t}\|_2^2$ and $\eta_{{\bf f}_t}\ge\eta_{\bf f}$ gives $\tilde L_{\bf f}({\bf f}_{t+1};\bar{\bf w}_t)-\tilde L_{\bf f}({\bf f}_{t};\bar{\bf w}_t)\ge\sigma\eta_{\bf f}m_{\bf f}\|{\bf g}_{{\bf f}_t}\|_2^2$. Since ${\bf f}_t,{\bf f}_{t+1}\in\mathcal S_{\bf f}$ and $\tilde L_{\bf f}$ coincides with $L(\cdot,\bar{\bf w}_t)$ on $\mathcal S_{\bf f}$, this is \eqref{eq:armijo_ascent}.
Replacing $(m_{{\bf f}},M_{{\bf f}},{\bf g}_{{\bf f}_t},{\bf d}_{{\bf f}_t})$ by $(m_{\bar{\bf w}},M_{\bar{\bf w}},\nabla_{\bar{\bf w}}\tilde L_{\bar{\bf w}},{\bf P}_{\bar{\bf w}}\nabla_{\bar{\bf w}}\tilde L_{\bar{\bf w}})$ and using \eqref{eq:zr_def_ppga} and \eqref{eq:armijo_w_ppga} in place of \eqref{eq:zf_def_ppga} and \eqref{eq:armijo_f_ppga} yields the same result for the $\bar{\bf w}$-block with $\bar\eta_{\bar{\bf w}}$.
\end{proof}
Lemma~\ref{lem:bt_term_ppga} combines two complementary guarantees: the Armijo condition in \eqref{eq:armijo_f_ppga} ensures an ascent proportional to $\eta_{{\bf f}_t}\|\nabla_{\bf f}\tilde L_{\bf f}({\bf f}_t)\|_2^2$, while the backtracking rule keeps the accepted step bounded away from zero, $\eta_{{\bf f}_t}\ge\eta_{\bf f}>0$. Neither alone suffices, since a mere ascent condition could be met by vanishing steps and a fixed step could violate the ascent condition. Note that the step bound $\bar\eta_{\bf f}$ depends on $L_g$ and the preconditioner only through $m_{\bf f}/M_{\bf f}^2$, so that a well-conditioned $\tilde{\bf B}({\bf f})^{-1}$ admits larger accepted steps, and the backtracking rule attains them without knowledge of these constants.

\subsection{Fast Alternating Convergent GPI}
\label{subsec:fac}
Although Lemma~\ref{lem:bt_term_ppga} guarantees that the AC-GPI ascends monotonically, repeatedly refining one block until inner convergence incurs unnecessary computations, since its solution changes once the other block is updated. Moreover, Lemma~\ref{lem:bt_term_ppga} shows that a single Armijo-accepted step already yields a uniform ascent proportional to the squared block gradient, so that running the inner loop to convergence is not needed for monotone ascent. Inspired by this observation, we propose the fast alternating convergent GPI (FAC-GPI), which performs only one Armijo-accepted step for the satellite block ${\bf f}$ and one for the BS block $\bar{\bf w}$ in each round, so that each block immediately accounts for the latest update of the other block. Each round of the FAC-GPI thus costs two Armijo-accepted GPI steps, whereas a round of the AC-GPI costs $2\bar I_{\rm in}$ such steps on average, where $\bar I_{\rm in}$ denotes the average number of inner steps per block, which is finite by Corollary~\ref{cor:inner_bound}; whether this per-round saving translates into a lower total cost depends on the number of rounds, which we examine through the rate bounds in Section~\ref{sec:conv} and the numerical results in Section~\ref{sec:num}.
By Lemma~\ref{lem:bt_term_ppga}, every accepted step increases \eqref{eq:Lagrangian_v1}, so this one-sweep structure preserves monotonic ascent without over-solving either block in an inner loop.
The FAC-GPI is summarized in Algorithm~\ref{alg:Precoding_shortcut}.

\begin{algorithm}[t]
\caption{FAC-GPI}\label{alg:Precoding_shortcut}
{\bf{initialize}}: $\mathbf f_0$, $\bar{\mathbf w}_0$

Set the iteration count $t = 0$

\While{$\delta_t>\epsilon$}{
  Build $\tilde{\mathbf A}(\mathbf f_t)$ and $\tilde{\mathbf B}(\mathbf f_t)$ from \eqref{A(w)} and \eqref{B(w)}

  Choose $\eta\in\{\eta_0,\beta\eta_0,\beta^2\eta_0,\ldots\}$ satisfying \eqref{eq:armijo_f_ppga}

  Compute $\mathbf f_{t+1} =  \mathbf f_t+\eta\,\tilde{\mathbf B}^{-1}(\mathbf f_t)\nabla_{\mathbf f}\tilde L_{\mathbf f}(\mathbf f_t)$

    Normalize $\mathbf f_{t+1} = \mathbf f_{t+1}/\|\mathbf f_{t+1}\|$

    Build $\tilde{\mathbf C}(\bar{\mathbf w}_t)$ and $\tilde{\mathbf D}(\bar{\mathbf w}_t)$ from \eqref{eq:CKKT} and \eqref{eq:DKKT}

  Choose $\eta\in\{\eta_0,\beta\eta_0,\beta^2\eta_0,\ldots\}$ satisfying \eqref{eq:armijo_w_ppga}

  Compute $\bar{\mathbf w}_{t+1} = \bar{\mathbf w}_t+\eta\,\tilde{\mathbf D}^{-1}(\bar{\mathbf w}_t)\nabla_{\bar{\mathbf w}}\tilde L_{\bar{\mathbf w}}(\bar{\mathbf w}_t)$

    Normalize $\bar{\mathbf w}_{t+1} = \bar{\mathbf w}_{t+1}/\|\bar{\mathbf w}_{t+1}\|$

    $\delta_t\leftarrow$ $\|\mathbf f_{t+1}-\mathbf f_t\| + \|\bar{\mathbf w}_{t+1}-\bar{\mathbf w}_t\|$

  $t\leftarrow t+1$
}
\Return $\mathbf f_t$, $\bar{\mathbf w}_t$
\end{algorithm}

\subsection{Backhaul-Control Update and Overall Algorithm}
\label{subsec:backhaul_control}
We next address the update of the multiplier $\mu$, which balances sum SE maximization against the self-backhaul constraint in \eqref{eq:Lagrangian_v1}: a large $\mu$ enforces the constraint at the expense of the sum SE, whereas a small $\mu$ favors the sum SE while possibly violating the constraint.
To find an appropriate $\mu$, we present a  bisection based method.
The bisection is justified by the fact that, since $L$ is affine in $\mu$, the constraint slack $R^{\sf es}-\sum_{k=1}^{K}R_k^{\sf ue}$ evaluated at a maximizer of $L(\cdot,\mu)$ is nondecreasing in $\mu$ \cite{boyd2004convex}, so that the feasibility of the solution changes at most once along $\mu$. 
Although the AC-GPI and FAC-GPI return stationary points rather than global maximizers, the numerical results in Section~\ref{sec:num} demonstrate the effectiveness of the bisection-based update of $\mu$.

We maintain the search interval $[\mu_{\sf lo},\mu_{\sf hi}]$ of $\mu$, where $\mu_{\sf lo}$ and $\mu_{\sf hi}$ are initialized such that the interval contains a proper $\mu$.
For a given $\mu$, once the stationary solutions ${\bf f}$ and $\Bar{\mathbf{w}}$ are obtained, we examine whether the self-backhaul constraint is satisfied.
If the constraint is satisfied, smaller value of $\mu$ is used to identify a stationary point with a reduced constraint penalty.
In this case, the upper end of the interval is updated as $\mu_{\sf hi}\leftarrow\mu$.
Conversely, if the constraint is violated, we increase $\mu$ by updating the lower end of the interval as $\mu_{\sf lo}\leftarrow\mu$.
Then, $\mu$ is updated as $\mu\leftarrow(\mu_{\sf lo}+\mu_{\sf hi})/2$, and this procedure is repeated until $\mu_{\sf hi}-\mu_{\sf lo}\le\epsilon_0$.
Combining the step-wise precoding-direction updates with the bisection update of $\mu$ yields the overall FAC-GPI algorithm summarized in Algorithm~\ref{alg:main_QRS_WSM_notau}.

\begin{algorithm}[t]
\caption{Overall FAC-GPI}
\label{alg:main_QRS_WSM_notau}
{\bf{initialize}}: ${\bf f}_{0}$, $\bar{\bf w}_{0}$, $\mu_{\sf lo}$, and $\mu_{\sf hi}$\\
Set  $t_0\leftarrow0$\\
\While{$\mu_{\sf hi}-\mu_{\sf lo}>\epsilon_0$}{
Set $\mu_{t_0}\leftarrow(\mu_{\sf lo}+\mu_{\sf hi})/2$ and $t_0\leftarrow t_0+1$\\
Update ${\bf f}_{t_0}$ and $\bar{\bf w}_{t_0}$ using Algorithm~\ref{alg:Precoding_shortcut} with fixed $\mu_{t_0}$\\
\eIf{$\sum_{k = 1}^{K} { R}_{k}^{\sf{ue}}\left(\mathbf{{ f}}, \mathbf{{\bar w}}\right) \!\leq\!  R^{\sf es}\left({\mathbf{f}},\bar{\mathbf{w}}\right)$}{
$\mu_{\sf hi}\leftarrow\mu_{t_0}$
}{
$\mu_{\sf lo}\leftarrow\mu_{t_0}$
}
}
\Return ${\bf f}_{t_0}$ and $\bar{\bf w}_{t_0}$
\end{algorithm}

\subsection{Complexity Analysis and Low-Complexity Implementation}
\label{subsec:cr}
Let us first analyze the computational complexity of the FAC-GPI in Algorithm~\ref{alg:Precoding_shortcut}.
The main computational burden for the algorithm is on the inversion of ${\bf{B}}_{\sf KKT}({\bf f})$ and ${\bf{D}}_{\sf KKT}(\bar {\bf{w}})$.
Since the complexity of the inverse of a matrix of size $M \times M$ is $\mathcal{O}(M^3)$, the complexity of computing ${\bf B}_{\sf KKT}^{-1}({{\bf f}})$ is $\mathcal{O} (M^3)$.
To compute the precoder $\bar {\bf{w}}$, the inversion of an $N \times N$ matrix requires  $\mathcal{O}(N^3)$.
Since ${\bf D}_{\sf KKT}(\bar {\bf{w}})$ consists of $K$ such sub-matrices, the overall complexity of evaluating ${\bf D}_{\sf KKT}^{-1}({\bar{{\bf w}}})$ is $\mathcal{O} (N^3K)$.
Furthermore, as ${\bf f}$ and $\Bar{\mathbf{w}}$ are updated alternately, the total complexity of Algorithm~\ref{alg:Precoding_shortcut} is computed as $\mathcal{O}(T_{\sf GPI}\mathrm{max}(M^3, KN^3))$, where $T_{\sf GPI}$  denotes the number of outer-loop iterations.
With the bisection update of $\mu$, the overall  complexity in Algorithm~\ref{alg:main_QRS_WSM_notau} is $\mathcal{O}\!\left(T_{\mu}T_{\sf GPI}\max(M^3,KN^3)\right)$,
where $T_{\mu}$ denotes the number of bisection iterations.
To further reduce the computational burden, we introduce a low-complexity implementation of the proposed GPI method, referred to as proposed (Low), following the SM-based approach in \cite{yoo2026scalable}.
Since the scalar factor $\lambda_{\sf den}$ does not affect the GPI update, it is omitted.

Using ${\bf S}^{\sf sat}=\frac{(\sigma^{\sf sat})^2}{P_s}{\bf I}_M$ and ${\bf U}^{\sf bs}_k=\frac{P_s}{P_t}{\bf g}^{\sf int}_k({\bf g}^{\sf int}_k)^{\sf H}$, the satellite-side KKT matrix in \eqref{B(w)} is a scaled identity plus $K$ rank-one terms:
\begin{align}
{\bf B}_{\sf KKT}({\bf f})={\bf B}^{\sf sat}_{\sf diag}+\sum_{k=1}^{K}c^{\sf su}_{k}{\bf g}^{\sf int}_k({\bf g}^{\sf int}_k)^{\sf H},
\label{eq:sat_B_struct}
\end{align}
where $c^{\sf su}_{k}=\frac{P_s}{P_t}\frac{1-\mu}{{\bf f}^{\sf H}({\bf U}^{\sf bs}_k+c_k^{\sf S}{\bf I}_M){\bf f}}$ and ${\bf B}^{\sf sat}_{\sf diag}=\big(\frac{\mu((\sigma^{\sf sat})^2/P_s+c)}{{\bf f}^{\sf H}({\bf S}^{\sf sat}+c{\bf I}_M){\bf f}}+\sum_{k=1}^{K}\frac{(1-\mu)c_k^{\sf S}}{{\bf f}^{\sf H}({\bf U}^{\sf bs}_k+c_k^{\sf S}{\bf I}_M){\bf f}}\big){\bf I}_M$.
Starting from the scaled identity ${\bf B}^{\sf sat}_{\sf diag}$, whose inverse is immediate, ${\bf B}_{\sf KKT}^{-1}({\bf f})$ is obtained exactly by applying the SM formula $K$ times, once per rank-one term, with $\mathcal{O}(KM^2)$ operations instead of $\mathcal{O}(M^3)$.

For the BS update, the $k'$th $N\times N$ diagonal block of the KKT matrix in \eqref{eq:DKKT} can be written as
\begin{align}
{\bf D}_{{\sf KKT},k'}^{\sf sub}={\bf D}^{\sf sh}-c_{k'}^{\sf ue}{\bf h}_{k'}{\bf h}_{k'}^{\sf H},
\label{eq:sub-matrix_of_Bkkt_approximated}
\end{align}
where ${\bf D}^{\sf sh}={\bf D}_{\sf diag}^{\sf sub}+c^{\sf be}{\bf h}^{\sf int}({\bf h}^{\sf int})^{\sf H}+\sum_{k=1}^{K}c_{k}^{\sf ue}{\bf h}_{k}{\bf h}_{k}^{\sf H}$ is shared by all $K$ blocks, ${\bf D}_{\sf diag}^{\sf sub}=\big(\frac{\mu d^{\sf S}}{\bar{\bf w}^{\sf H}({\bf U}^{\sf sat}+d^{\sf S}{\bf I}_{NK})\bar{\bf w}}+\sum_{k=1}^{K}c_{k}^{\sf ue}\big(\frac{(\sigma^{\sf bs})^2}{P_t}+d_k\big)\big){\bf I}_N$, $c_{k}^{\sf ue}=\frac{1-\mu}{\bar{\bf w}^{\sf H}({\bf S}_{k}^{\sf bs}+d_k{\bf I}_{NK})\bar{\bf w}}$, and $c^{\sf be}=\frac{\mu P_t}{P_s\,\bar{\bf w}^{\sf H}({\bf U}^{\sf sat}+d^{\sf S}{\bf I}_{NK})\bar{\bf w}}$.
Applying the SM formula to each block separately would require $\mathcal O(K^2N^2)$ operations, since each block contains $K+1$ rank-one terms. Instead, $({\bf D}^{\sf sh})^{-1}$ is computed once by $K+1$ SM updates from the scaled identity ${\bf D}_{\sf diag}^{\sf sub}$, and the inverse of each block follows from a single rank-one SM downdate, so that all $K$ block inverses are obtained with $\mathcal O(KN^2)$ operations instead of $\mathcal O(KN^3)$ for direct inversion. Compared with \cite{yoo2026scalable}, the shared matrix ${\bf D}^{\sf sh}$ additionally contains the rank-one term $c^{\sf be}{\bf h}^{\sf int}({\bf h}^{\sf int})^{\sf H}$ from the BS-to-ES interference, which is absorbed into the shared SM updates at no additional order of complexity.

Accordingly, the low-complexity implementation computes the GPI update exactly and reduces the per-iteration complexity from $\mathcal{O}(\max(M^3,KN^3))$ to $\mathcal{O}(K\max(M^2,N^2))$.

\section{Convergence Analysis}
\label{sec:conv}
In this section, we analyze the convergence of the FAC-GPI and AC-GPI for a fixed $\mu$.
We first establish convergence to stationary points, and then characterize the global stationarity rate.

\subsection{Convergence to Stationary Points}


Lemma~\ref{lem:bt_term_ppga} guarantees sufficient ascent for each accepted step, but it does not by itself imply the convergence of the alternating scheme: the guarantee is given for one block with the other block fixed, so that the partial gradients controlled by Lemma~\ref{lem:bt_term_ppga} are evaluated at different iterates. To close this gap, we show in this subsection that the alternating updates jointly drive the iterates to a stationary point of \eqref{eq:Lagrangian_v1} on $\mathcal S_{\bf f}\times\mathcal S_{\bar{\bf w}}$.

We first show that each inner loop of the AC-GPI terminates within a number of steps bounded uniformly over rounds.
\begin{corollary}[Finite inner loops]\label{cor:inner_bound}
For any $\epsilon_{\rm in}>0$, each inner loop of Algorithm~\ref{alg:Precoding_conv} terminates after at most
\begin{align}
I(\epsilon_{\rm in})=1+\left\lfloor\frac{4\eta_0^2\bar M^2\Delta_L}{\sigma\underline{c}\,\epsilon_{\rm in}^2}\right\rfloor
\label{eq:inner_bound}
\end{align}
inner steps in every round, where $\Delta_L=\max L-\min L$ over $\mathcal S_{\bf f}\times\mathcal S_{\bar{\bf w}}$, $\bar M=\max\{M_{\bf f},M_{\bar{\bf w}}\}$, and $\underline{c}=\min\{\eta_{\bf f}m_{\bf f},\eta_{\bar{\bf w}}m_{\bar{\bf w}}\}$.
\end{corollary}
\begin{proof}
Consider the ${\bf f}$-phase of a round; the $\bar{\bf w}$-phase is identical. Every inner step that does not terminate the loop satisfies
\begin{align}
    \label{eq:col_eq1}
\epsilon_{\rm in}<\|{\bf f}^{(s+1)}-{\bf f}^{(s)}\|_2\le2\eta_0M_{\bf f}\|\nabla_{\bf f}\tilde L_{\bf f}({\bf f}^{(s)};\bar{\bf w}_t)\|_2,
\end{align}
where the last inequality follows from $\|{\bf y}/\|{\bf y}\|_2-{\bf f}\|_2\le2\|{\bf y}-{\bf f}\|_2$ for $\|{\bf f}\|_2=1$, $\eta\le\eta_0$, and Assumption~\ref{ass:precond_ppga}. By \eqref{eq:armijo_ascent} and \eqref{eq:col_eq1}, each such step increases $L$ by more than $\sigma\eta_{\bf f}m_{\bf f}\epsilon_{\rm in}^2/(4\eta_0^2M_{\bf f}^2)$. Since $L$ is continuous on the compact set $\mathcal S_{\bf f}\times\mathcal S_{\bar{\bf w}}$, the total increase within the phase is at most $\Delta_L<\infty$, so that the number of such steps is at most $4\eta_0^2M_{\bf f}^2\Delta_L/(\sigma\eta_{\bf f}m_{\bf f}\epsilon_{\rm in}^2)$. Adding the final step yields \eqref{eq:inner_bound}.
\end{proof}
We now establish the convergence of the AC-GPI and the FAC-GPI.
\begin{theorem}[Convergence of the AC-GPI and FAC-GPI]
\label{thm:conv_ppga}
For fixed $\mu$, let the iterates be generated by the FAC-GPI in Algorithm~\ref{alg:Precoding_shortcut} or by the AC-GPI in Algorithm~\ref{alg:Precoding_conv} with $\epsilon_{\rm in}>0$. Then:
\begin{enumerate}
\item 
$L({\bf f}_{t+1},\bar{\bf w}_{t+1})\!\ge\! L({\bf f}_{t},\bar{\bf w}_{t})$ for all $t$.
\item 
$\|\nabla_{\bf f}\tilde L_{\bf f}({\bf f}_{t};\bar{\bf w}_{t})\|_2\to0$ and
$\|\nabla_{\bar{\bf w}}\tilde L_{\bar{\bf w}}(\bar{\bf w}_{t};{\bf f}_{t+1})\|_2\to0$.
\item 
$\|{\bf f}_{t+1}-{\bf f}_t\|_2\to0$ and $\|\bar{\bf w}_{t+1}-\bar{\bf w}_t\|_2\to0$.
\item Every accumulation point of the iterates is a first-order stationary point of $L$ on
$\mathcal{S}_{\bf f}\times\mathcal{S}_{\bar{\bf w}}$.
\end{enumerate}
\end{theorem}
\begin{proof}
We first prove the result for the FAC-GPI and then extend it to the AC-GPI.

\emph{Step 1: Monotonic ascent.}
The $\mathbf f$-block Armijo condition gives $
\tilde L_{\bf f}({\bf f}_{t+1})
\ge
\tilde L_{\bf f}({\bf f}_{t}).$
Since $\tilde L_{\bf f}$ coincides with $L$ on $\mathcal S_{\bf f}$ for fixed $\bar{\bf w}_t$, this implies
\begin{align}
L({\bf f}_{t+1},\bar{\bf w}_{t})
\ge
L({\bf f}_{t},\bar{\bf w}_{t}).
\end{align}
Likewise, the $\bar{\bf w}$-block update gives
\begin{align}
L({\bf f}_{t+1},\bar{\bf w}_{t+1})
\ge
L({\bf f}_{t+1},\bar{\bf w}_{t}).
\end{align}
Combining the two inequalities proves monotonic ascent.

\emph{Step 2: Objective convergence.}
The feasible set $\mathcal{S}_{\bf f}\times\mathcal{S}_{\bar{\bf w}}$ is compact and $L$ is continuous.
Thus, the non-decreasing objective sequence is bounded above and converges.

\emph{Step 3: Vanishing block gradients.}
By Lemma~\ref{lem:bt_term_ppga}, each accepted $\mathbf f$-block step satisfies \eqref{eq:armijo_ascent} with $\eta_{{\bf f}_t}\ge\eta_{\bf f}>0$. By Step 1, for every $t$, we further have
\begin{align}
L({\bf f}_{t+1},\bar{\bf w}_{t+1})-L({\bf f}_{t},\bar{\bf w}_{t})
&\ge\sigma\eta_{\bf f}m_{\bf f}\|\nabla_{\bf f}\tilde L_{\bf f}({\bf f}_t;\bar{\bf w}_t)\|_2^2.
\label{eq:round_ascent}
\end{align}
Summing over $t$ and telescoping the left-hand side with $L({\bf f}_{T},\bar{\bf w}_{T}) \leq L^\infty$ yields
\begin{align}
\sigma\eta_{\bf f}m_{\bf f}\sum_{t=0}^{T-1}\|\nabla_{\bf f}\tilde L_{\bf f}({\bf f}_t;\bar{\bf w}_t)\|_2^2
&\le L^\infty-L({\bf f}_{0},\bar{\bf w}_{0})<\infty,
\end{align}
where $L^\infty$ denotes the limit of the objective sequence in Step 2. Letting $T\to\infty$ gives $\sum_{t}\|\nabla_{\bf f}\tilde L_{\bf f}({\bf f}_t;\bar{\bf w}_t)\|_2^2<\infty$, and hence $\|\nabla_{\bf f}\tilde L_{\bf f}({\bf f}_t;\bar{\bf w}_t)\|_2\to0$. Similarly, applying the $\bar{\bf w}$-block counterpart of Lemma~\ref{lem:bt_term_ppga}, which starts from $(\bar{\bf w}_t;{\bf f}_{t+1})$, yields $\|\nabla_{\bar{\bf w}}\tilde L_{\bar{\bf w}}(\bar{\bf w}_t;{\bf f}_{t+1})\|_2\to0$.

\textit{Step 4: Stationarity of accumulation points.}
Since the feasible set $\mathcal{S}_{\bf f}\times\mathcal{S}_{\bar{\bf w}}$ is compact, the sequence $\{({\bf f}_t,\bar{\bf w}_t)\}$ has at least one accumulation point; let $({\bf f}_{t_j},\bar{\bf w}_{t_j})\to({\bf f}^\circ,\bar{\bf w}^\circ)$.
The ${\bf f}$-block update in \eqref{eq:zf_def_ppga} satisfies
\begin{align}
\nonumber
\|{\bf f}_{t+1}-{\bf f}_t\|_2
&\!=\!
\left\|
\frac{{\bf f}_t+\eta_{{\bf f}_t}{\bf d}_{{\bf f}_t}}
{\|{\bf f}_t+\eta_{{\bf f}_t}{\bf d}_{{\bf f}_t}\|_2}
-{\bf f}_t
\right\|_2\!\overset{(a)}{\le}
2\eta_{{\bf f}_t}\|{\bf d}_{{\bf f}_t}\|_2
\overset{(b)}{\le}
2\eta_0\|{\bf d}_{{\bf f}_t}\|_2
\nonumber\\
&\overset{(c)}{\le}
2\eta_0M_{\bf f}
\big\|\nabla_{\bf f}\tilde L_{\bf f}
({\bf f}_t;\bar{\bf w}_t)\big\|_2
\overset{(d)}{\rightarrow}0,
\label{eq:step_vanish}
\end{align}
where (a) follows from
$\|{\bf y}/\|{\bf y}\|_2-{\bf f}\|_2
\le 2\|{\bf y}-{\bf f}\|_2$
for any $\|{\bf f}\|_2=1$ and ${\bf y}\neq{\bf 0}$,
(b) holds because the backtracking rule only shrinks the initial step size so that
$\eta_{{\bf f}_t}\le\eta_0$,
(c) follows from the uniform preconditioner bound in
Assumption~\ref{ass:precond_ppga},
and (d) follows from Step 3.
Applying the same argument to the $\bar{\bf w}$-block yields $\|\bar{\bf w}_{t+1}-\bar{\bf w}_t\|_2\to0$.
By Step 3 and the joint continuity of $\nabla_{\bf f}\tilde L_{\bf f}$ in $({\bf f},\bar{\bf w})$:
\begin{align}
\nabla_{\bf f}\tilde L_{\bf f}
({\bf f}_{t_j};\bar{\bf w}_{t_j})
\rightarrow
\nabla_{\bf f}\tilde L_{\bf f}
({\bf f}^{\circ};\bar{\bf w}^{\circ})
={\bf 0}.
\end{align}
Moreover, since
$\|{\bf f}_{t+1}-{\bf f}_{t}\|_2\rightarrow0$,
we have
${\bf f}_{t_j+1}\rightarrow{\bf f}^{\circ}$.
The $\bar{\bf w}$-block gradient is evaluated using the updated satellite precoder ${\bf f}_{t_j+1}$, and Step 3 gives $\nabla_{\bar{\bf w}}\tilde L_{\bar{\bf w}}
(\bar{\bf w}_{t_j};{\bf f}_{t_j+1})
\rightarrow {\bf 0}.$
Hence, by the joint continuity of $\nabla_{\bar{\bf w}}\tilde L_{\bar{\bf w}}$ in $(\bar{\bf w},{\bf f})$, $\nabla_{\bar{\bf w}}\tilde L_{\bar{\bf w}}
(\bar{\bf w}^{\circ};{\bf f}^{\circ})
={\bf 0}.$
Therefore, both block gradients vanish at
$({\bf f}^{\circ},\bar{\bf w}^{\circ})$, and thus
$({\bf f}^{\circ},\bar{\bf w}^{\circ})$
is a first-order stationary point of
\eqref{eq:Lagrangian_v1} on
$\mathcal{S}_{\bf f}\times\mathcal{S}_{\bar{\bf w}}$.

In the AC-GPI, every inner update is an Armijo-accepted step, so that Lemma~\ref{lem:bt_term_ppga} applies to each inner step, and Steps 1 and 2 hold unchanged.
For Step 3, the $\mathbf f$-phase of round $t$ performs $I_t\le I(\epsilon_{\rm in})$ inner steps with ${\bf f}^{(0)}={\bf f}_t$ and ${\bf f}^{(I_t)}={\bf f}_{t+1}$, where ${\bf g}^{(s)}_{\bf f}=\nabla_{\bf f}\tilde L_{\bf f}({\bf f}^{(s)};\bar{\bf w}_t)$.
Summing \eqref{eq:armijo_ascent} over the inner steps gives
\begin{align}
L({\bf f}_{t+1},\bar{\bf w}_{t})-L({\bf f}_{t},\bar{\bf w}_{t})
&\ge
\sigma\eta_{\bf f}m_{\bf f}\sum_{s=0}^{I_t-1}\|{\bf g}^{(s)}_{\bf f}\|_2^2
\nonumber\\
&\ge
\sigma\eta_{\bf f}m_{\bf f}\|\nabla_{\bf f}\tilde L_{\bf f}({\bf f}_t;\bar{\bf w}_t)\|_2^2,
\label{eq:bw_phase_ascent}
\end{align}
where the last inequality keeps only the first term ${\bf g}^{(0)}_{\bf f}=\nabla_{\bf f}\tilde L_{\bf f}({\bf f}_t;\bar{\bf w}_t)$.
Since the $\bar{\bf w}$-phase likewise starts from $(\bar{\bf w}_t;{\bf f}_{t+1})$, the same summation yields its counterpart for the $\bar{\bf w}$-block. Hence, \eqref{eq:bw_phase_ascent} reduces the AC-GPI to the same per-round ascent bound \eqref{eq:round_ascent} as the FAC-GPI, and the remaining argument of Step 3 applies without modification, so that both block gradients vanish.
For Step 4, \eqref{eq:step_vanish} bounds only a single step, whereas one round of the AC-GPI consists of $I_t$ inner steps.
Applying the step bound in \eqref{eq:step_vanish} to each inner step gives $\|{\bf f}^{(s+1)}-{\bf f}^{(s)}\|_2\le2\eta_0M_{\bf f}\|{\bf g}^{(s)}_{\bf f}\|_2$. Hence,
\begin{align}
\|{\bf f}_{t+1}-{\bf f}_t\|_2
&\overset{(a)}{\le}\sum_{s=0}^{I_t-1}\|{\bf f}^{(s+1)}-{\bf f}^{(s)}\|_2
\le2\eta_0M_{\bf f}\sum_{s=0}^{I_t-1}\|{\bf g}^{(s)}_{\bf f}\|_2\nonumber\\
&\overset{(b)}{\le}2\eta_0M_{\bf f}\sqrt{I(\epsilon_{\rm in})\textstyle\sum_{s=0}^{I_t-1}\|{\bf g}^{(s)}_{\bf f}\|_2^2}\nonumber\\
&\overset{(c)}{\le}2\eta_0M_{\bf f}\sqrt{\frac{I(\epsilon_{\rm in})\big(L({\bf f}_{t+1},\bar{\bf w}_{t})-L({\bf f}_{t},\bar{\bf w}_{t})\big)}{\sigma\eta_{\bf f}m_{\bf f}}}
\rightarrow0,
\end{align}
where (a) follows from the triangle inequality applied to ${\bf f}_{t+1}-{\bf f}_t=\sum_{s=0}^{I_t-1}({\bf f}^{(s+1)}-{\bf f}^{(s)})$, (b) follows from the Cauchy-Schwarz inequality and $I_t\le I(\epsilon_{\rm in})$, and (c) follows from the first inequality in \eqref{eq:bw_phase_ascent}. 
The convergence follows since the per-round ascent $L({\bf f}_{t+1},\bar{\bf w}_{t})-L({\bf f}_{t},\bar{\bf w}_{t})\ge0$ is summable over $t$ by Step 2, and thus vanishes.
The same argument applies to the $\bar{\bf w}$-block, and the remaining part of Step 4 is unchanged.

\end{proof}

\begin{remark}[Overall algorithm]
\label{rem:overall}
The bisection in Algorithm~\ref{alg:main_QRS_WSM_notau} halves the interval $[\mu_{\sf lo},\mu_{\sf hi}]$ at each outer iteration, so that it terminates in at most $\lceil\log_2(\Delta_0/\epsilon_0)\rceil$ outer iterations, where $\Delta_0$ denotes the initial length of the search interval.
Within each outer iteration, Algorithm~\ref{alg:Precoding_shortcut} runs with fixed $\mu$ and terminates after finitely many rounds, since the change in the precoders between consecutive rounds converges to zero by Theorem~\ref{thm:conv_ppga} and each round requires finitely many backtracking steps by Lemma~\ref{lem:bt_term_ppga}.
Hence, Algorithm~\ref{alg:main_QRS_WSM_notau} terminates after finitely many accepted GPI steps.
\end{remark}

\subsection{Global Stationarity Rate}
\label{subsec:rates}
Theorem~\ref{thm:conv_ppga} established that every accumulation point of the iterates is first-order stationary.
We now quantify the convergence through a global stationarity rate.
We first derive the per-round estimates in terms of the  gradient  and then obtain the global stationarity rates of both variants.
Write
$\mathbf{x}_t=(\mathbf{f}_t,\bar{\mathbf{w}}_t)$,
$\mathbf{g}^{t}_{\mathbf{f}}=\nabla_{\mathbf{f}}\tilde{L}_{\mathbf{f}}(\mathbf{f}_t;\bar{\mathbf{w}}_t)$,
and
$\mathbf{g}^{t}_{\bar{\mathbf{w}}}=\nabla_{\bar{\mathbf{w}}}\tilde{L}_{\bar{\mathbf{w}}}(\bar{\mathbf{w}}_t;\mathbf{f}_{t+1})$.
Let $\eta_{\mathbf{f}}=\min\{\eta_0,\beta\bar\eta_{\mathbf{f}}\}$  and 
$\kappa=\sigma\min\{\eta_{\mathbf{f}}m_{\mathbf{f}},\eta_{\bar{\mathbf{w}}}m_{\bar{\mathbf{w}}}\}$.
We also define $
C_1=2\eta_0\max\{M_{\mathbf{f}},M_{\bar{\mathbf{w}}}\}$ and $
C_2(I)=\max\bigl\{2,\ 1+2L_g^2C_1^2I\bigr\}.$

\begin{lemma}[Per-round ascent]\label{lem:perround}
For the FAC-GPI, the following holds for all $t$.
\begin{align}
L(\mathbf{x}_{t+1})-L(\mathbf{x}_t)&\ge \kappa'\|\operatorname{grad}L(\mathbf{x}_t)\|^2,\label{eq:ascent}
\end{align}
where $\kappa'=\tfrac{\kappa}{C_2(1)}$.
For the AC-GPI, \eqref{eq:ascent} holds with $\kappa'$ replaced by $\kappa'_{\mathrm{bl}}=\kappa/C_2(I(\epsilon_{\mathrm{in}}))\le \kappa'$, where $I(\epsilon_{\mathrm{in}})$ is given in \eqref{eq:inner_bound}.
\end{lemma}

\begin{proof}
For each block,
\begin{align}
&A_{\mathbf{f}}=L(\mathbf{f}_{t+1},\bar{\mathbf{w}}_t)-L(\mathbf{x}_t)
\overset{(a)}{\ge} \sigma\eta_{\mathbf{f}_t}m_{\mathbf{f}}\|\mathbf{g}^{t}_{\mathbf{f}}\|^2
\overset{(b)}{\ge} \kappa\|\mathbf{g}^{t}_{\mathbf{f}}\|^2,\label{eq:blockascent}\\
&A_{\bar{\mathbf{w}}}=L(\mathbf{x}_{t+1})-L(\mathbf{f}_{t+1},\bar{\mathbf{w}}_t)
\ge \kappa\|\mathbf{g}^{t}_{\bar{\mathbf{w}}}\|^2,
\end{align}
where (a) follows from the Armijo condition \eqref{eq:armijo_f_ppga}, which the accepted step of Lemma~\ref{lem:bt_term_ppga} satisfies, and Assumption~\ref{ass:precond_ppga}, and (b) uses the step lower bound $\eta_{{\bf f}_t}\ge\eta_{\bf f}$ in Lemma~\ref{lem:bt_term_ppga} and the definition of $\kappa$; the $\bar{\mathbf{w}}$-block follows likewise.
Then, we have 
\begin{align}
    \label{eq:lemma4_eq1}
    L(\mathbf{x}_{t+1})-L(\mathbf{x}_t)=A_{\mathbf{f}}+A_{\bar{\mathbf{w}}}
\ge\kappa(\|\mathbf{g}^{t}_{\mathbf{f}}\|^2+\|\mathbf{g}^{t}_{\bar{\mathbf{w}}}\|^2).
\end{align}
Since ${\bf f}_t$ is a unit vector, normalizing any ${\bf y}\neq{\bf 0}$
obeys $\left\|
\frac{\bf y}{\|{\bf y}\|_2}-{\bf f}_t\right\|_2
\le2\|{\bf y}-{\bf f}_t\|_2$.
Substituting the accepted trial point ${\bf y}={\bf f}_t+\eta_{{\bf f}_t}{\bf P}_{\bf f}{\bf g}_{\bf f}^t,$
so that ${\bf f}_{t+1}={\bf y}/\|{\bf y}\|_2$, gives
\begin{align}
\|{\bf f}_{t+1}-{\bf f}_t\|_2
\overset{(a)}{\le}
2\eta_{{\bf f}_t}
\|{\bf P}_{\bf f}{\bf g}_{\bf f}^t\|_2
\overset{(b)}{\le}
2\eta_0 M_{\bf f}\|{\bf g}_{\bf f}^t\|_2\le
C_1\|{\bf g}_{\bf f}^t\|_2,
\label{eq:stepbd}
\end{align}
where $(a)$ applies the normalization bound above,  $(b)$ uses the accepted-step upper bound $\eta_{{\bf f}_t}\le \eta_0$ and the preconditioner bound in \eqref{eq:bounding_ppga}, and
$C_1=2\eta_0\max\{M_{\mathbf{f}},M_{\bar{\mathbf{w}}}\}$.
The full gradient at $\mathbf{x}_t$ is
$\operatorname{grad}L(\mathbf{x}_t)=(\mathbf{g}^{t}_{\mathbf{f}},\bar{\mathbf{g}}^{\,t}_{\bar{\mathbf{w}}})$,
where $\bar{\mathbf{g}}^{\,t}_{\bar{\mathbf{w}}}=\nabla_{\bar{\mathbf{w}}}\tilde{L}_{\bar{\mathbf{w}}}(\bar{\mathbf{w}}_t;\mathbf{f}_t)$
is evaluated at the $t$th iterate $\mathbf{f}_t$, whereas the accepted update uses
$\mathbf{g}^{t}_{\bar{\mathbf{w}}}=\nabla_{\bar{\mathbf{w}}}\tilde{L}_{\bar{\mathbf{w}}}(\bar{\mathbf{w}}_t;\mathbf{f}_{t+1})$.
Since $\bar{\mathbf{g}}^{\,t}_{\bar{\mathbf{w}}}$ and $\mathbf{g}^{t}_{\bar{\mathbf{w}}}$ share the same $\bar{\mathbf{w}}_t$ but are evaluated at $\mathbf{f}_t$ and $\mathbf{f}_{t+1}$, respectively, we have the mismatch bound
\begin{align}
\|\bar{\mathbf{g}}^{\,t}_{\bar{\mathbf{w}}}-\mathbf{g}^{t}_{\bar{\mathbf{w}}}\|
\overset{(a)}{\le} L_g\|\mathbf{f}_{t+1}-\mathbf{f}_t\|
\overset{(b)}{\le} L_gC_1\|\mathbf{g}^{t}_{\mathbf{f}}\|_2,
\label{eq:mismatch}
\end{align}
where $(a)$ is the $L_g$-Lipschitz continuity of the gradient and $(b)$ is~\eqref{eq:stepbd}.
By the triangle inequality and \eqref{eq:mismatch}, we have $\|\bar{\mathbf{g}}^{\,t}_{\bar{\mathbf{w}}}\|\le\|\mathbf{g}^{t}_{\bar{\mathbf{w}}}\|+L_gC_1\|\mathbf{g}^{t}_{\mathbf{f}}\|$, which further yields
\begin{align}
\nonumber
&\|\!\operatorname{grad}L(\mathbf{x}_t)\!\|^2
\!=\!\|\mathbf{g}^{t}_{\mathbf{f}}\|^2+\|\bar{\mathbf{g}}^{\,t}_{\bar{\mathbf{w}}}\|^2\!
\overset{(a)}{\le}\!\|\mathbf{g}^{t}_{\mathbf{f}}\|^2\!+\!2\|\mathbf{g}^{t}_{\bar{\mathbf{w}}}\|^2\!+\!2L_g^2C_1^2\|\mathbf{g}^{t}_{\mathbf{f}}\|^2\\
&{=}(1+2L_g^2C_1^2)\|\mathbf{g}^{t}_{\mathbf{f}}\|^2+2\|\mathbf{g}^{t}_{\bar{\mathbf{w}}}\|^2\!
\overset{(b)}{\le} \!C_2(1)\bigl(\|\mathbf{g}^{t}_{\mathbf{f}}\|^2+\|\mathbf{g}^{t}_{\bar{\mathbf{w}}}\|^2\bigr),
\label{eq:bounding}
\end{align}
where $(a)$ squares this bound using $(u+v)^2\le2u^2+2v^2$, and $(b)$ dominates both coefficients by $C_2(1)=\max\{2,\,1+2L_g^2C_1^2\}$.
Then, \eqref{eq:lemma4_eq1} is further lower bounded as
\begin{align}
L(\mathbf{x}_{t+1})-L(\mathbf{x}_t)
{\ge}\kappa\bigl(\|\mathbf{g}^{t}_{\mathbf{f}}\|^2+\|\mathbf{g}^{t}_{\bar{\mathbf{w}}}\|^2\bigr)
&\overset{(a)}{\ge}\frac{\kappa}{C_2(1)}\,\|\operatorname{grad}L(\mathbf{x}_t)\|^2
\nonumber
\\
&=\kappa'\|\operatorname{grad}L(\mathbf{x}_t)\|^2,
\end{align}
where $\kappa'=\kappa/C_2(1)$.
Here, $(a)$ is  from \eqref{eq:bounding}.

In the AC-GPI, the ${\mathbf{f}}$-phase of round $t$ performs $I_t\le I(\epsilon_{\mathrm{in}})$ inner steps with ${\mathbf{f}}^{(0)}={\mathbf{f}}_t$ and ${\mathbf{f}}^{(I_t)}={\mathbf{f}}_{t+1}$.
The $\bar{\mathbf{w}}$-phase likewise runs an inner loop.
Let $\mathbf{g}^{(s)}_{{\mathbf{f}}}=\nabla_{{\mathbf{f}}}\tilde{L}_{{\mathbf{f}}}({\mathbf{f}}^{(s)};\bar{\mathbf{w}}_t)$ denote the gradient at the $s$th inner step. 
The ${\mathbf{f}}$-phase ascent is then lower-bounded as
\begin{align}
A_{{\mathbf{f}}}
\overset{(a)}{\ge}\kappa\sum_{s=0}^{I_t-1}\|\mathbf{g}^{(s)}_{{\mathbf{f}}}\|^2
\overset{(b)}{\ge}\kappa\|\mathbf{g}^{t}_{{\mathbf{f}}}\|^2,
\label{eq:bxascent}
\end{align}
where $(a)$ applies \eqref{eq:blockascent} to each inner step $s=0,\dots,I_t-1$ and telescopes the left-hand side to $A_{\mathbf{f}}$, and $(b)$ keeps only the first term $\mathbf{g}^{(0)}_{{\mathbf{f}}}=\mathbf{g}^{t}_{{\mathbf{f}}}$.
We further have
\begin{align}
&\|\mathbf{f}_{t+1}-\mathbf{f}_t\|
\nonumber\\
&\overset{(a)}{\le\!} C_1\!\sum_{s=0}^{I_t-1}\|\mathbf{g}_{{\mathbf{f}}}^{(s)}\|\!
\overset{(b)}{\le} \!C_1\sqrt{\textstyle I_t\!\sum_{s=0}^{I_t-1}\|\mathbf{g}_{{\mathbf{f}}}^{(s)}\|^2}
\overset{(c)}{\le} C_1\sqrt{I(\epsilon_{\mathrm{in}})A_{\mathbf{f}}/\kappa},
\label{eq:drift}
\end{align}
where $(a)$ is~\eqref{eq:stepbd} applied stepwise with the triangle inequality,
$(b)$ is Cauchy-Schwarz inequality, and $(c)$
uses $I_t\le I(\epsilon_{\mathrm{in}})$ from Corollary~\ref{cor:inner_bound} with  \eqref{eq:bxascent}.
By the triangle inequality and the first inequality in \eqref{eq:mismatch}, $\|\bar{\mathbf{g}}^{\,t}_{\bar{\mathbf{w}}}\|\le\|\mathbf{g}^{t}_{\bar{\mathbf{w}}}\|+L_g\|\mathbf{f}_{t+1}-\mathbf{f}_t\|$. Squaring this bound using $(u+v)^2\le2u^2+2v^2$ and inserting~\eqref{eq:drift} gives
\begin{align}
\|\bar{\mathbf{g}}^{\,t}_{\bar{\mathbf{w}}}\|^2
\le 2\|\mathbf{g}^{t}_{\bar{\mathbf{w}}}\|^2
+2L_g^2C_1^2 I(\epsilon_{\mathrm{in}})A_{\mathbf{f}}/\kappa.
\label{eq:bxmismatch}
\end{align}
Combining these bounds, $\|\operatorname{grad}L(\mathbf{x}_t)\|$ is upper bounded by the per-round ascent as
\begin{align}
\nonumber
\|\!\operatorname{grad}L(\mathbf{x}_t)\|^2&\overset{(a)}{\le}\!\!\frac{(1+2L_g^2C_1^2 I(\epsilon_{\mathrm{in}}))A_{\mathbf{f}}+2A_{\bar{\mathbf{w}}}}{\kappa}
\\
&\!\overset{(b)}{\le}\!\frac{C_2(I(\epsilon_{\mathrm{in}}))}{\kappa}\bigl(A_{\mathbf{f}}+A_{\bar{\mathbf{w}}}\bigr),
\label{eq:bxnormequiv}
\end{align}
where $(a)$ inserts~\eqref{eq:bxmismatch} into
$\|\operatorname{grad}L(\mathbf{x}_t)\|^2=\|\mathbf{g}^{t}_{\mathbf{f}}\|^2+\|\bar{\mathbf{g}}^{\,t}_{\bar{\mathbf{w}}}\|^2$
and uses $\|\mathbf{g}^{t}_{\mathbf{f}}\|^2\le A_{\mathbf{f}}/\kappa$ in \eqref{eq:bxascent} and similarly, $\|\mathbf{g}^{t}_{\bar{\mathbf{w}}}\|^2\le A_{\bar{\mathbf{w}}}/\kappa$, and $(b)$ bounds
both coefficients by $C_2(I(\epsilon_{\mathrm{in}}))=\max\{2,\,1+2L_g^2C_1^2 I(\epsilon_{\mathrm{in}})\}$.
Since $A_{\mathbf{f}}+A_{\bar{\mathbf{w}}}=L(\mathbf{x}_{t+1})-L(\mathbf{x}_t)$,
\eqref{eq:bxnormequiv} is~\eqref{eq:ascent} with $\kappa'$ replaced by
$\kappa'_{\mathrm{bl}}=\frac{\kappa}{C_2(I(\epsilon_{\mathrm{in}}))}\le\kappa'$.
\end{proof}

Compared with Lemma~\ref{lem:bt_term_ppga}, which guarantees ascent for a single block step in terms of the partial gradient evaluated at the current point of that block, Lemma~\ref{lem:perround} measures the ascent of an entire round in terms of the gradient $\operatorname{grad}L(\mathbf{x}_t)$ evaluated at a single iterate. The price of this conversion is the factor $C_2(\cdot)$ in $\kappa'$, which accounts for the mismatch between the block gradients used by the alternating updates and $\operatorname{grad}L(\mathbf{x}_t)$. Since \eqref{eq:ascent} lower-bounds the per-round ascent by the squared gradient norm at the same iterate, summing it over the rounds directly yields the following global stationarity rate.

\begin{corollary}[Global convergence rates]\label{cor:global}
For the FAC-GPI with every $T\ge1$, the following holds:
\begin{align}
\min_{0\le t<T}\|\operatorname{grad}L(\mathbf{x}_t)\|^2
\le \frac{L^\infty-L_0}{\kappa'\,T},
\label{eq:global}
\end{align}
where $L_0=L(\mathbf{x}_0)$, and the same bound with $\kappa'$ replaced by $\kappa'_{\mathrm{bl}}$ for the
AC-GPI.
\end{corollary}

\begin{proof}
Telescoping \eqref{eq:ascent}, $L^\infty-L_0  \ge \kappa'\sum_{t=0}^{T-1}\|\operatorname{grad}L(\mathbf{x}_t)\|^2
\ge \kappa'\,T\min_{0\le t<T}\|\operatorname{grad}L(\mathbf{x}_t)\|^2.$
\end{proof}
To compare the two variants at equal cost, let $S$ be the total number of  GPI steps, each costing one KKT inversion.
Since the FAC-GPI and AC-GPI perform $2$ and at most $2I(\epsilon_{\mathrm{in}})$ such steps per round, respectively, $S$ GPI steps correspond to at least $T=S/2$ and $T=S/(2I(\epsilon_{\mathrm{in}}))$ rounds for the FAC-GPI and AC-GPI, respectively. Substituting these into \eqref{eq:global} and using $L^\infty-L_0\le\Delta_L$ with $\Delta_L$ in Corollary~\ref{cor:inner_bound} gives
\begin{align}
\min_{0\le t<S/2}\|\operatorname{grad}L(\mathbf{x}_t)\|^2&\le\frac{2C_2(1)\Delta_L}{\kappa S},\nonumber\\
\min_{0\le t<S/(2I(\epsilon_{\mathrm{in}}))}\|\operatorname{grad}L(\mathbf{x}_t)\|^2&\le\frac{2I(\epsilon_{\mathrm{in}})C_2(I(\epsilon_{\mathrm{in}}))\Delta_L}{\kappa S},
\label{eq:global_S}
\end{align}
so that the AC-GPI bound is weaker by the factor $I(\epsilon_{\mathrm{in}})C_2(I(\epsilon_{\mathrm{in}}))/C_2(1)$, where $I(\epsilon_{\mathrm{in}})$ reflects the higher per-round cost of the AC-GPI and $C_2(I(\epsilon_{\mathrm{in}}))/C_2(1)\ge1$ reflects its weaker per-round guarantee. Note that this comparison concerns the guaranteed bounds rather than the actual convergence speed, since the factor $C_2(I(\epsilon_{\mathrm{in}}))$ stems from the mismatch bound \eqref{eq:drift} over the inner steps. Nevertheless, since \eqref{eq:global_S} provides a tighter guarantee on the gradient norm for the FAC-GPI at an equal number of GPI steps, it suggests that the FAC-GPI converges faster than the AC-GPI for a given computational budget, which is confirmed numerically in Section~\ref{sec:num}.

\section{Numerical Results} \label{sec:num}
We consider an integrated TN-NTN network in which the
satellite backhaul and terrestrial access links share the
$3.5$ GHz band, with cross-link interference managed through
joint precoding. 
The terrestrial network follows the 3GPP rural macrocell (RMa) scenario \cite{3gpp38901} with an inter-site distance of $5$ km.
The BS with $N=32$ antennas (8 dBi element gain, 35 m height)
serves $K=8$ single-antenna users (0 dBi antenna gain and 1.5 m height) uniformly distributed within the cell, with a
minimum BS-user distance of 35 m.
The BS-user channels follow the RMa
LoS/NLoS path-loss models with distance-dependent LoS
probabilities and shadow fading \cite{3gpp38901},
while their spatial correlation follows the one-ring model
\cite{adhi:tit:13}. 

The LEO satellite at an altitude of $600$ km and an
elevation angle of $60^\circ$ employs an $8\times8$
UPA ($M=64$, 5 dBi element gain) and transmits at
$P_s=55$ dBm over a $30$ MHz bandwidth, corresponding
to an EIRP density of approximately $33$ dBW/MHz
\cite{3gpp38821}. Since the ES and users are nearly
co-directional as viewed from the satellite, their
channels are modeled as ${\bf g}=\sqrt{\beta^{\sf es}}\,\nu^{\sf es}
{\bf a}(\theta^{\sf es},\phi^{\sf es})$ and ${\bf g}_k^{\sf int}=\sqrt{\beta_k^{\sf sat}}\,\nu_k
{\bf a}(\theta_k,\phi_k)$,
where ${\bf a}(\cdot)$ denotes the array response and $\nu^{\sf es}$ and $\nu_k$ denote scalar 
fading coefficients.
The Rician factors are set to $\kappa^{\sf es}=20$ dB and $\kappa_k=10$ dB for LoS users. 
The satellite-user LoS probability is $0.94$, with a $20$ dB clutter loss for NLoS links and a $3$ dB polarization loss \cite{3gpp38811}.
The ES, located $1.5$ km from the BS at a height of $10$ m, employs a $1$ m dish with a $30$ dBi boresight gain pointed at the satellite, so that its gain toward the BS is -10 dBi \cite{ITU-R-S465}, and the BS-ES link is LoS. 
All results are averaged over $1000$ independent
channel realizations.

The considered baseline methods are as follows:
\begin{itemize}
   \item {\bf{ISTS-SCA}}: The successive convex approximation (SCA)-based beamforming algorithm jointly optimizes the satellite and BS beamformers using the S-procedure and penalty functions \cite{wang2021joint}.
   The algorithm has a computational complexity of $\mathcal{O}(T_{\sf SCA}(4K+5)^4\mathrm{max}{\{N,M\}}^{1/2}\log{(1+\delta}))$, where $T_{\sf SCA}$ denotes the number of outer iterations and $\delta$ represents the solution accuracy.
   \item {{\bf WMMSE}: The WMMSE algorithm \cite{christensen2008weighted}, which jointly optimizes the satellite and BS precoders.}
    \item {\bf{RZF}}: The conventional linear regularized zero-forcing (RZF), where the satellite and BS precoders are formed from the available channels.
\end{itemize}
Note that the proposed AC-GPI and FAC-GPI schemes, along with RZF, are evaluated under full-power transmission, whereas the other baseline methods are implemented without enforcing full-power transmission.

\begin{figure}[!t]\centering
{\resizebox{0.85\columnwidth}{!}{\includegraphics{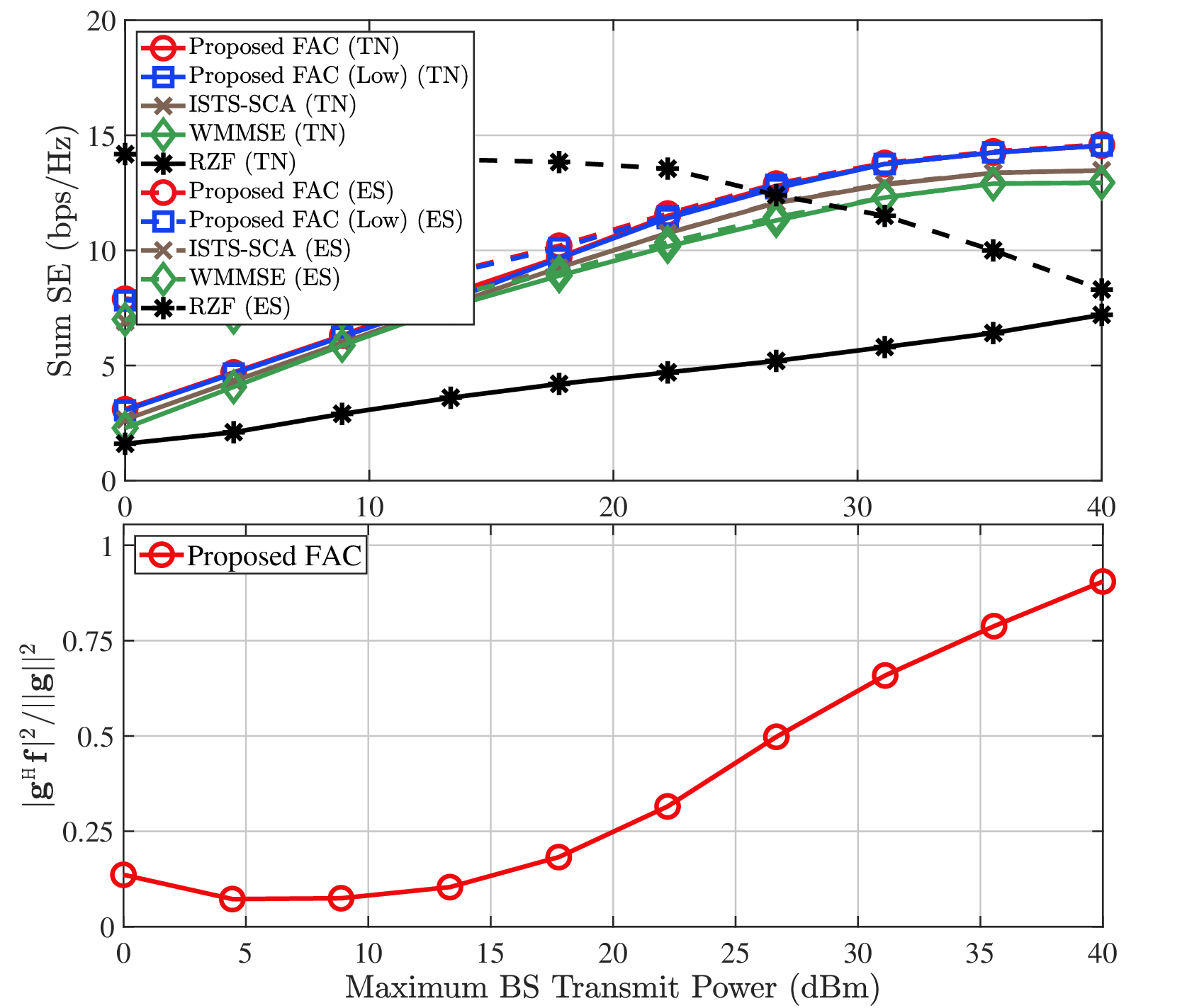}}}
\caption{
 	 The user sum SE (solid lines) and ES SE (dashed lines) (top), and the  effective satellite gain of the proposed FAC-GPI (bottom), versus the maximum BS transmit power.}  
 	\label{fig:WSM-weight0.3}
\end{figure}

\subsection{SE versus BS Transmit Power}
In Fig.~\ref{fig:WSM-weight0.3}, we plot the user sum SE and ES SE (top) and the normalized effective satellite gain (bottom) versus the maximum BS transmit power.
As shown in Fig.~\ref{fig:WSM-weight0.3}, the proposed FAC-GPI achieves higher user sum SE than the WMMSE, ISTS-SCA, and RZF baselines, while the proposed (Low) achieves nearly identical SE.
At low BS transmit power, the user sum SE is power-limited, and the BS-to-ES interference is weak.
Since the ES and users are nearly co-directional as viewed from the satellite, optimizing $\mathbf f$ effectively controls the power delivered toward both the ES and users.
In this regime, FAC-GPI reduces the effective satellite gain toward the ES to mitigate satellite-to-user interference, despite full-power transmission.
As the BS transmit power increases, FAC-GPI increases the effective satellite gain while suppressing BS-to-ES interference through the BS precoder.
Then, the ES SE increases at higher BS transmit powers, and the user sum SE approaches the backhaul SE in the high-power regime.
In this regard, these results demonstrate the effectiveness of the proposed joint precoding in managing inter-system interference under the self-backhaul constraint.



\begin{figure}[!t]
    \centering
    \subfigure[]{
    \includegraphics[width=0.82\linewidth]{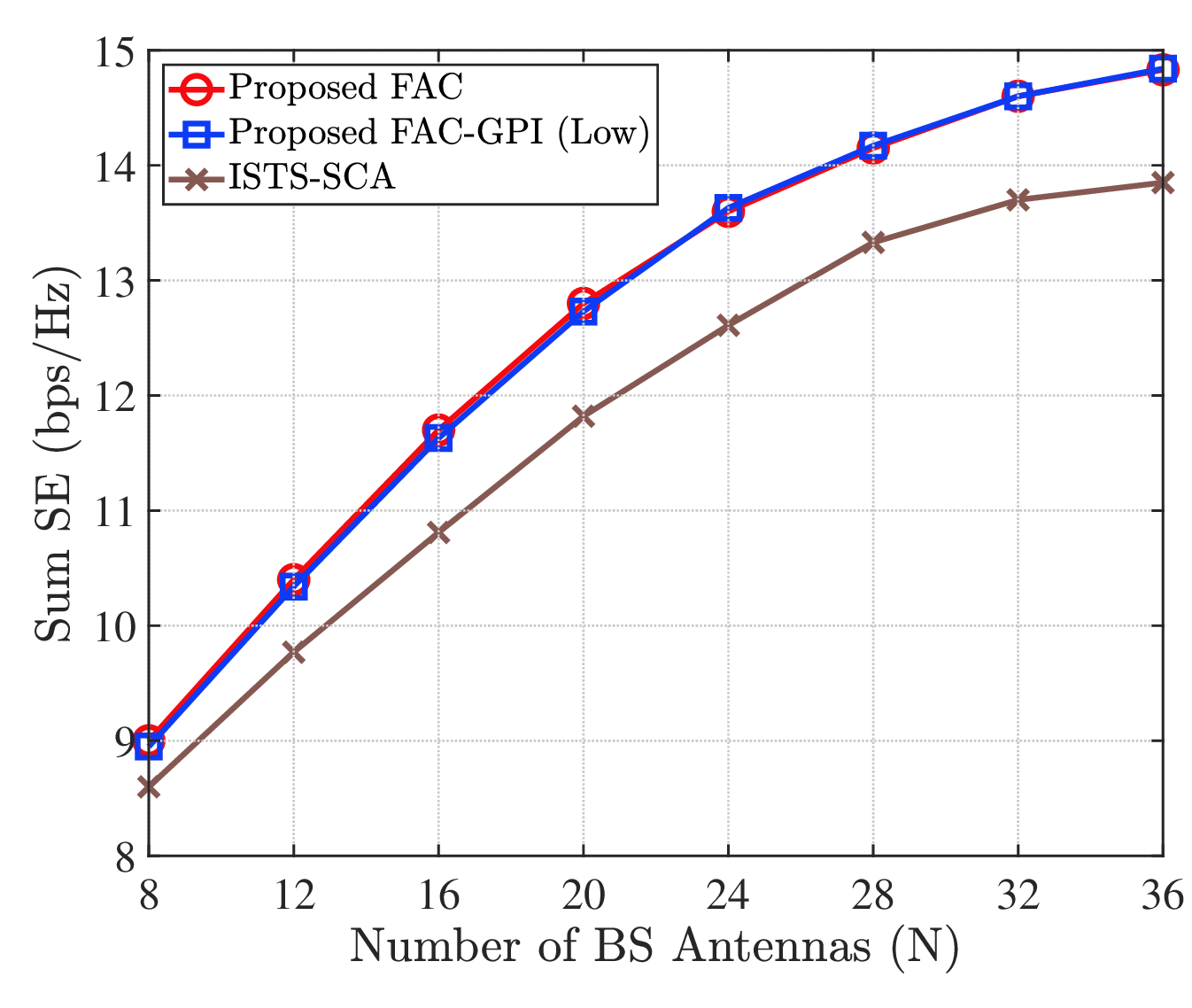}
    \label{fig:antenna_se}
    }
    \subfigure[]{
    \includegraphics[width=0.82\linewidth]{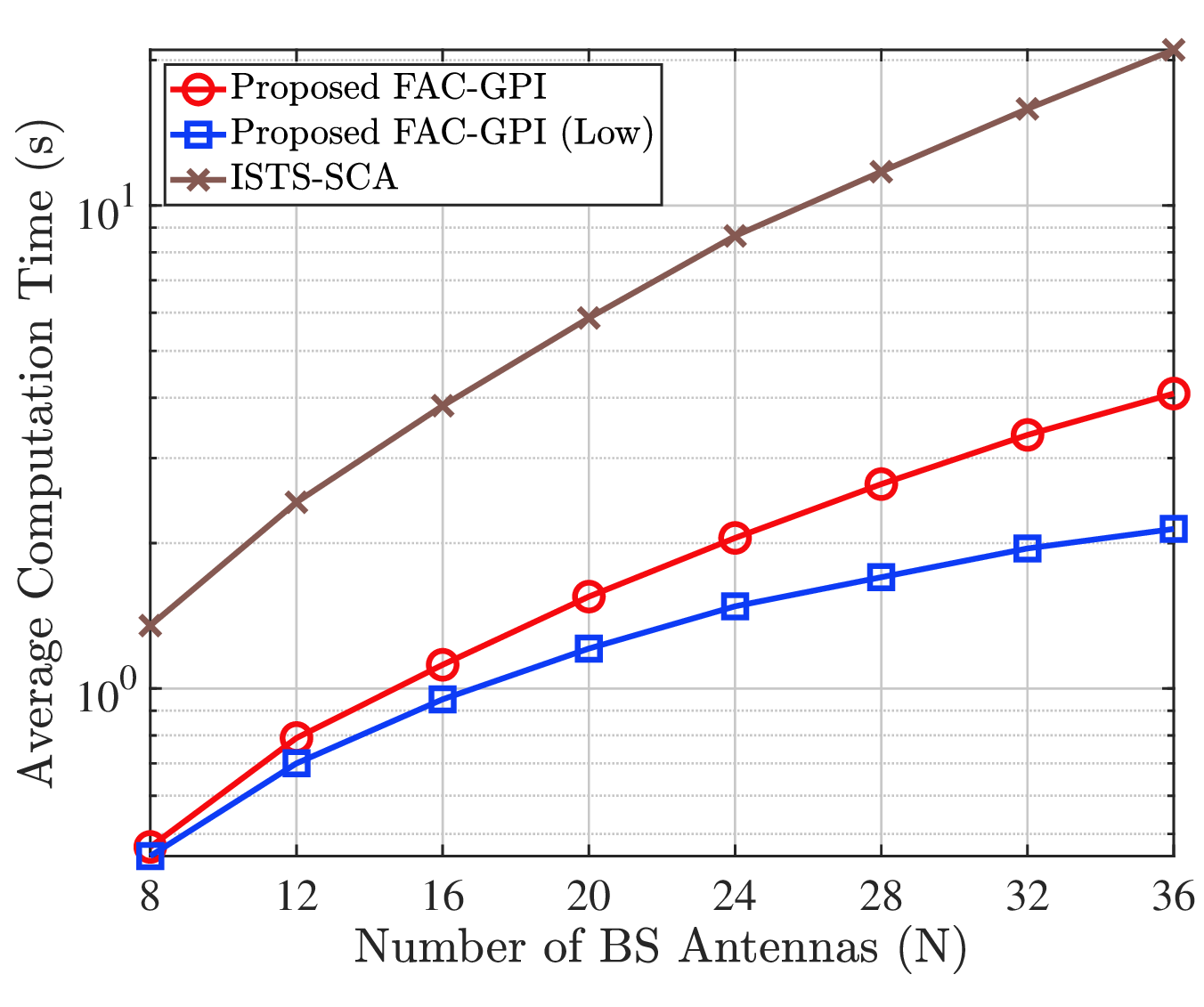}
    \label{fig:antenna_time}
    }
    \caption{The sum SE (top) and computation time (bottom) versus the number of BS antennas for   $P_t = 30$ dBm.}
    \label{fig:antena}
\end{figure}

\subsection{Effect of the Number of BS Antennas}
\label{subsec:sim_power}
Fig.~\ref{fig:antena} compares the proposed FAC-GPI algorithms with the ISTS-SCA baseline in terms of the sum SE and  average computation time as the number of BS antennas $N$ increases.
As shown in Fig.~\ref{fig:antena}(a), the proposed algorithm achieves the highest sum SE over the entire range of $N$, while the proposed (Low) algorithm maintains  the same performance.
Both proposed algorithms consistently outperform ISTS-SCA in terms of the sum SE.
Fig.~\ref{fig:antena}(b) shows that the computation time increases with $N$ for all methods, whereas the proposed algorithms require substantially less computation time than ISTS-SCA.
In particular, the proposed (Low) algorithm achieves the lowest computation time, and its computational advantage becomes more pronounced as $N$ increases.
These results show that the proposed (Low) algorithm achieves the same SE with the lowest computational complexity, making it well-suited for large-scale antenna systems.

\subsection{Convergence}
\label{subsec:sim_conv}
\begin{figure}[!t]
    \centering
    \subfigure[$P_t=0$ dBm]{
    \includegraphics[width=0.82\linewidth]{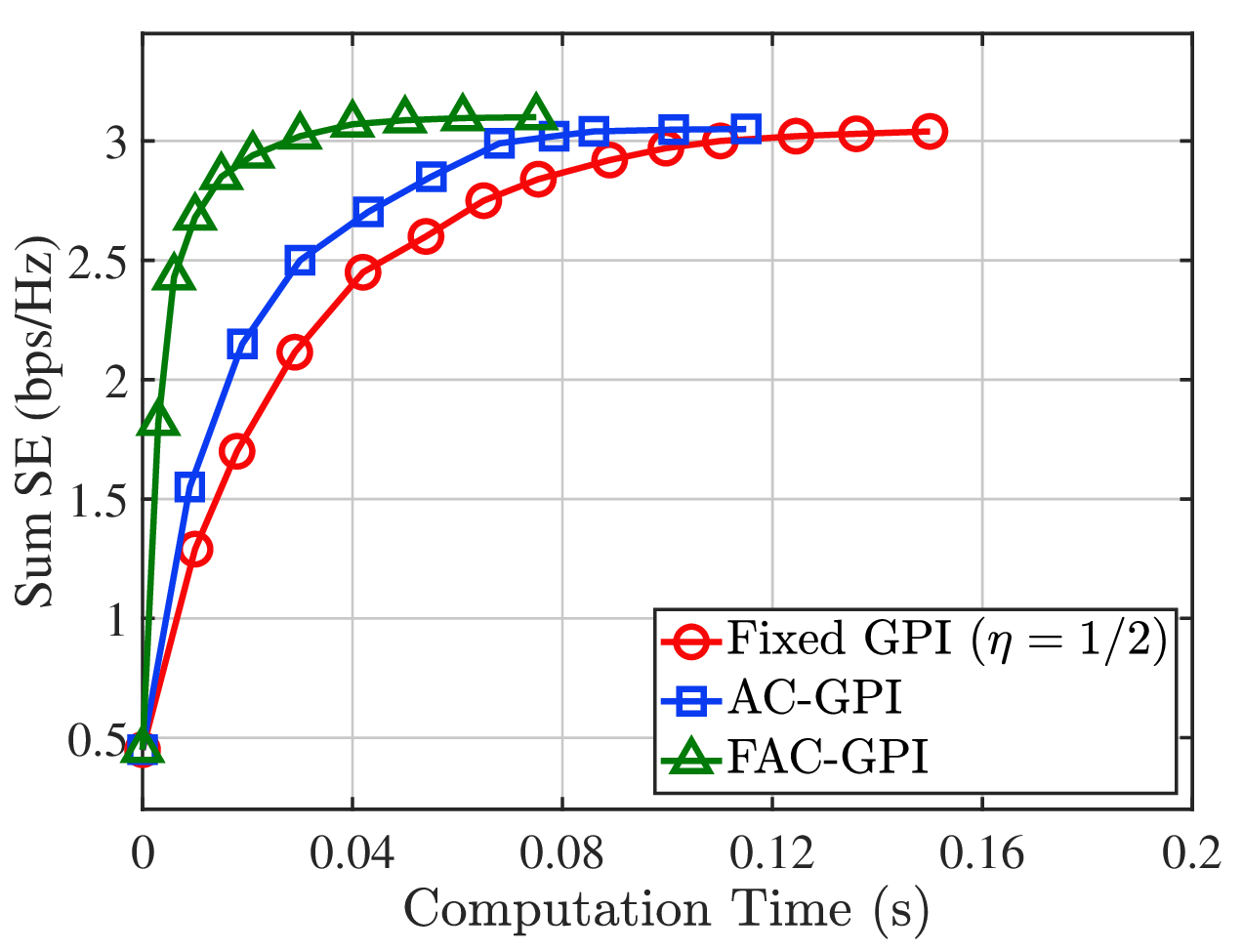}
    \label{fig:conv_P0dbf_perfect}
    }
    \subfigure[ $P_t=30$ dBm]{
    \includegraphics[width=0.82\linewidth]{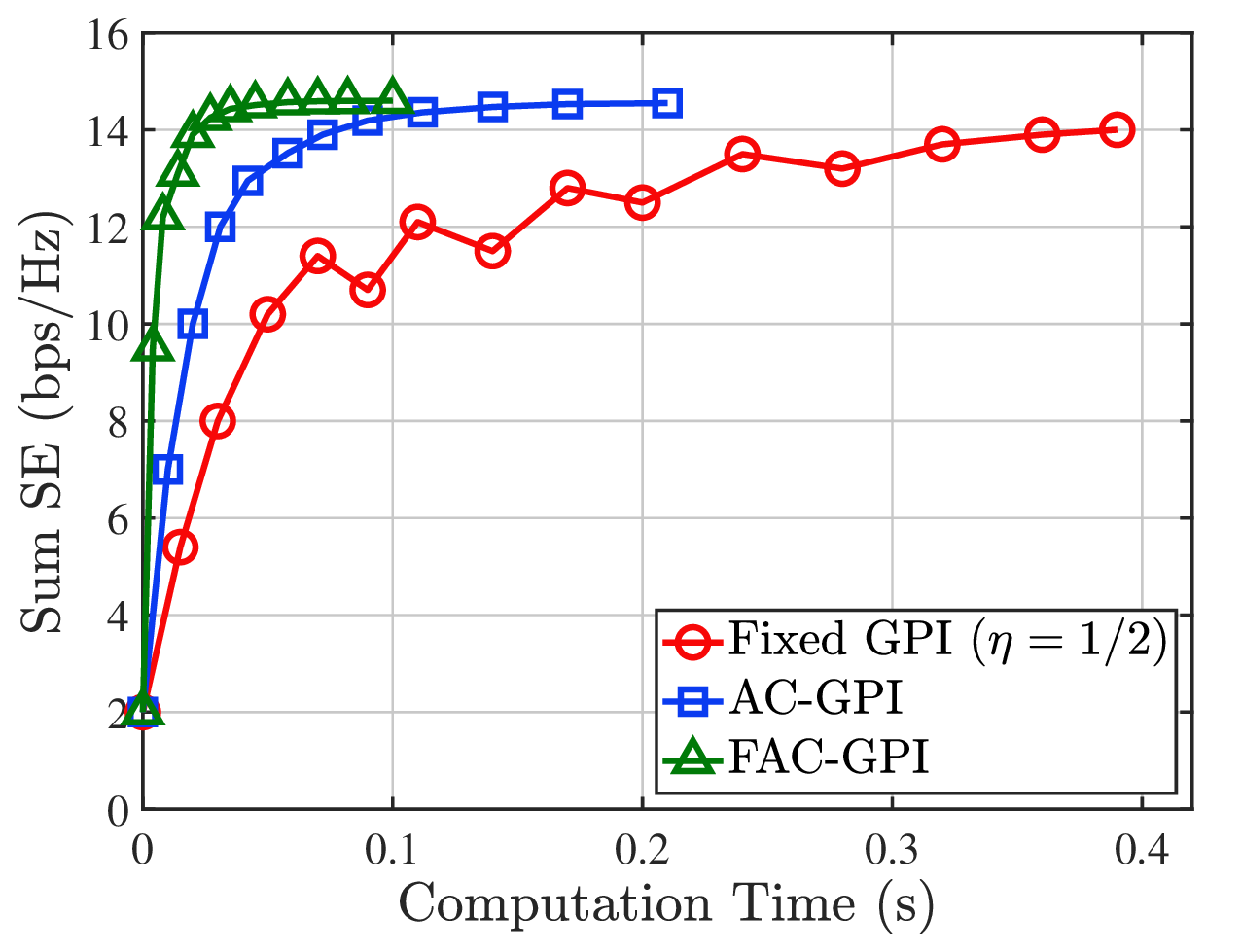}
    \label{fig:conv_P40dbf_perfect}
    }
            \subfigure[$P_t=30$ dBm]{
    \includegraphics[width=0.82\linewidth]{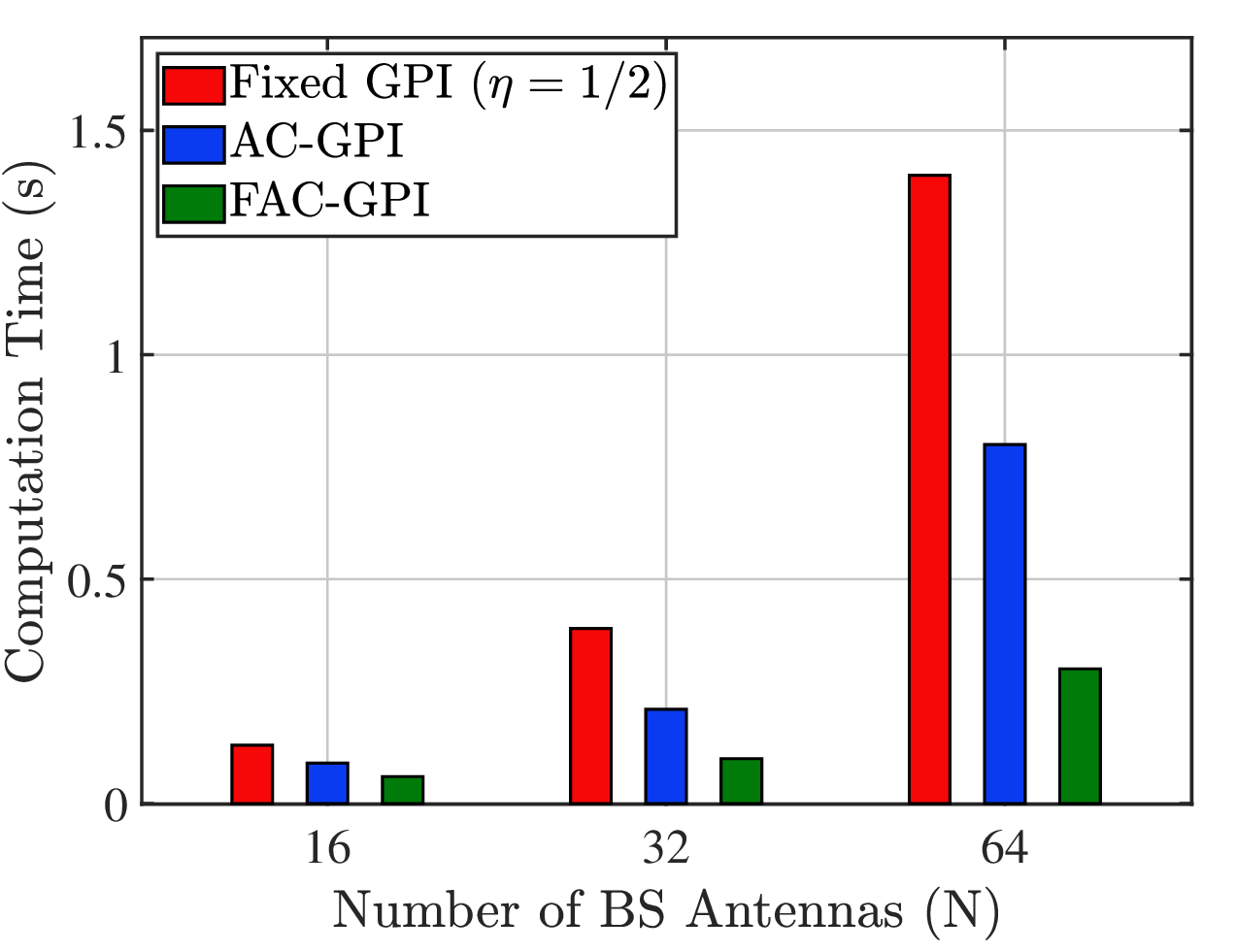}
    \label{fig:conv_antenna}
    }
    \caption{The sum SE versus computation time for (a) $P_t=0$ dBm and (b) $P_t=30$ dBm, and (c) the precoder computation time versus the number of BS antennas for $P_t=30$ dBm.}
    \label{fig:convergence}
\end{figure}

In Fig.~\ref{fig:convergence}, we compare the fixed-step alternating GPI ($\eta=1/2$) with the AC-GPI and FAC-GPI algorithm, both of which adjust $\eta$ via Armijo backtracking.
As shown in Fig.~\ref{fig:convergence}(a), in the low transmit-power regime, all three algorithms converge stably, while the AC-GPI and FAC-GPI converge faster than the fixed-step GPI.
In the high transmit-power regime of Fig.~\ref{fig:convergence}(b), the fixed-step GPI exhibits oscillatory convergence, since the steep and highly nonlinear objective yields a large Lipschitz constant that the fixed step may violate, causing overshooting.
In contrast, both the AC-GPI and FAC-GPI shrink the step to satisfy the sufficient-ascent condition and thus ascend monotonically.
Although the AC-GPI and FAC-GPI reach similar objective values, the FAC-GPI requires less computation time because it avoids repeated inner-loop updates while retaining a favorable per-round convergence bound.
Fig.~\ref{fig:convergence}(c) plots the precoder computation time versus the number of BS antennas $N$. 
As $N$ increases, the FAC-GPI achieves a larger computation-time advantage by avoiding repeated inner-loop updates.

\section{Conclusions}
In this paper, we proposed a fast and provably convergent precoding framework for maximizing the user sum SE in ISTS networks.
We formulated a user sum SE maximization problem subject to a self-backhaul constraint, which ensures that the user data forwarded by the BS do not exceed the data delivered to the ES.
We then reformulated the resulting non-convex problem into a GPI-friendly form and developed an alternating GPI-based precoding algorithm for the satellite and BS precoders.
By interpreting the GPI update as projected preconditioned gradient ascent, we replaced its fixed step with Armijo backtracking to obtain the AC-GPI, and, based on the per-step ascent guarantee, developed the FAC-GPI, which performs a single Armijo update for each precoding block per round.
For both algorithms, we established finite backtracking termination, monotone ascent, stationarity of accumulation points, and a global stationarity rate, and presented a low-complexity implementation that exploits the structure of the KKT matrices. Simulation results demonstrated that the FAC-GPI achieves higher SE than the baselines with fast and stable convergence.

\bibliographystyle{IEEEtran}


\bibliography{ref_dist_stin.bib}

\end{document}